\documentclass[jgrga]{AGUTeX}
\usepackage{color}
\usepackage{multirow}

\usepackage{color}
\usepackage[final]{graphicx}
\authorrunninghead{KITE ET AL.}

\titlerunninghead{LOCALIZED PRECIPITATION ON MARS}

\begin{document}

%
%

\title{Localized precipitation and runoff on Mars}
%

%
%



\authors{Edwin S. Kite, \altaffilmark{1,2}
Timothy I. Michaels, \altaffilmark{3} Scot Rafkin, \altaffilmark{3} Michael Manga, \altaffilmark{1,2} and William E. Dietrich. \altaffilmark{1}}

\altaffiltext{1}{Earth and Planetary Science, University of California Berkeley, Berkeley, California, USA.}

\altaffiltext{2}{Center for Integrative Planetary Science, University of California Berkeley, Berkeley, California, USA.}

\altaffiltext{3}{Department of Space Studies, Southwest Research Institute, Boulder, Colorado, USA.}



%
%


\begin{abstract}
We use the Mars Regional Atmospheric Modeling System (MRAMS) to simulate lake storms on Mars, finding that intense localized precipitation will occur for lake size $\geq$10$^3$ km$^2$. Mars has a low-density atmosphere, so deep convection can be triggered by small amounts of latent heat release. In our reference simulation, the buoyant plume lifts vapor above condensation level, forming a 20km-high optically-thick cloud. Ice grains grow to 200  $\mu$m radius and fall near (or in) the lake at mean rates up to 1.5 mm/hr water equivalent (maximum rates up to 6 mm/hr water equivalent). Because atmospheric temperatures outside the surface layer are always well below 273K, supersaturation and condensation begin at low altitudes above lakes on Mars. In contrast to Earth lake-effect storms, lake storms on Mars involve continuous precipitation, and their vertical velocities and plume heights exceed those of tropical thunderstorms on Earth. For lake sizes 10$^{2.5}$ - 10$^{3.5}$ km, plume vertical velocity scales linearly with lake area. Convection does not reach above the planetary boundary layer for lakes $\ll$10$^3$ km$^2$ or for atmospheric pressure $>$ O(10$^2$) mbar. Instead, vapor is advected downwind with little cloud formation. Precipitation occurs as snow, and the daytime radiative forcing at the land surface due to plume vapor and storm clouds is too small to melt snow directly ($<$ +10 W/m$^2$). However, if orbital conditions are favorable, then the snow may be seasonally unstable to melting and produce runoff to form channels. We calculate the probability of melting by running thermal models over all possible orbital conditions and weighting their outcomes by probabilities given by long-term integrations of the chaotic diffusion of solar system orbital elements \citep{las04}. With this approach, we determine that for an equatorial vapor source, sunlight 15\% fainter than at present, and snowpack with albedo 0.28 (0.35),  melting may occur with 4\% (0.1\%) probability. This rises to 56\% (12\%) if the ancient greenhouse effect was modestly (6K) greater than today.
\end{abstract}

%
%

%

\begin{article}

%
%

\section{Introduction}
Evidence for runoff on Mars shows it to be patchy in both space and time \citep{kra08,wil07,wei08,fas08b,hyn10,car00}, so perhaps past precipitation was also patchy. Because patchy surface vapor sources cannot persist in equilibrium with a dry atmosphere \citep{ric08,sot08}, vapor would have to be supplied from an environment not in equilibrium with surface conditions. Such environments can be transient, such as an impact lake, or long-lived, such as the base of a wet-based ice-sheet. They can be high-temperature, such as fumaroles (or a lava flow advancing over snowpack), or involve only moderate temperatures, such as groundwater discharge.

Here we use a mesoscale model to explore the atmospheric response to one example of a non-equilibrium vapor source: an ephemeral lake on a cold desert Mars. We track the fate of vapor supplied by the lake from release, through cloud formation, to precipitation, and consider whether the resulting snow will melt and provide runoff to form channels. Lake size, solar luminosity, lake geometry, and atmospheric pressure all affect the results. Only idealized results are presented: a companion paper (\citet{kit102}; henceforth Paper 2) uses the same model for a case study of the Juventae plateau inverted channel networks \citep{wei08}.

Low volumetric heat capacity makes the Mars atmosphere's response to lake vapor release similar to tropical moist convection on Earth, so we borrow ideas from tropical meteorology to understand our results (e.g. \citet{ema94}). Figure \ref{EQUIVTEMP} shows the low-pressure lake effect: condensation of a small amount of vapor in a thin atmosphere can produce strong convection, which in Earth's thick atmosphere would require condensation of a large amount of vapor and correspondingly high water surface temperatures.

Localized precipitation on a cold desert planet is normally transient precipitation. A warm, wet patch connected to the global atmosphere will lose water to cold traps elsewhere on the planet \citep{ric08}; the water table will withdraw to the subsurface because of evaporative losses \citep{sot08}. In the absence of an external heat source, evaporative and radiative cooling will cause any lake to quickly freeze \citep{lor05,con10}. (These arguments do not apply to springs, nor proglacial discharge of subglacial meltwater. In these cases, under cold conditions, any given parcel of water will freeze over, but a sustained vapor source can nevertheless exist at the discharge site.) As the ice thickens, ice surface temperature will fall and the lower saturation vapor pressure will cause the rate of vapor release to greatly decrease. For realistic external heat sources, the lake lifetime is still short. For example, consider an impact-generated lake near the freezing point overlying shocked basalt that is initially at 1000 $^\circ$C. The lake is assumed to be well-mixed by waves driven by lake-effect and impact-thermal storms, and convection driven by bottom heating. Icing-over is inevitable when the heat flow from the interior of the lake toward the surface is less than the evaporative and radiative losses at the surface. The maximum time before icing over, $t$, is therefore

\begin{equation}
t \approx \frac{D (T_{b} - T) c_{b} \rho_{b} }{ (E L_{vap} + \sigma T^4)}
\end{equation}


where $D$ is the depth of pervasive fracturing within the rock ejecta, $T_b$ = 1273K is the initial temperature of the basalt, $T$ = 278K is lake surface temperature, $c_{b}$ = 840 J/kg/K the specific heat capacity of the basalt, $\rho_{b}$ = 2000 kg/m$^3$ the density of the fractured basalt, $E$ = 2 kg/m$^2$/hr the evaporation rate, $L_{vap}$ = 2.5 x 10$^5$ J/kg the latent heat of vaporisation, and $\sigma$ = 5.67 x 10$^{-8}$ W/m$^2$/K$^4$ is the Stefan-Boltzmann constant. For $D$ = 100m we obtain $\sim$4 Earth years: a geological instant. The true timescale will be less. For example, if fractures within the ejecta anneal, the relevant timescale is conductive cooling of a half-space (the ejecta layer) by an isothermal boundary condition (the well-mixed lake) until the heat flow into the bottom of the lake is less than heat loss at the top of the lake \citep{tur02}:

\begin{equation}
t \approx \left( \frac{k (T_{b} - T) }{ E L_{vap} + \sigma T^4 } \right)^2 \frac{1}{ (\pi \kappa)}
\end{equation}

In this case, the conductive heat flow can only balance evaporative plus radiative losses for $\sim$3 days for thermal diffusivity $\kappa$ = 10$^{-6}$ m$^2$ s$^{-1}$: after this, an ice cover must form.

Therefore, we are interested in spatially restricted (10$^0$ - 10$^3$ km) water sources which cease to emit vapor in timescales $<$ 1 year. This is the domain of mesoscale modelling. We use the Mars Regional Atmospheric Modeling System (MRAMS), also used for entry, descent and landing simulations for the Mars Exploration Rovers, Mars Phoenix, and Mars Science Laboratory (Appendix A; \citet{raf01,mic08}). MRAMS explicitly resolves the size spectrum of dust and water ice aerosol for both cloud microphysics and radiative transfer, so it is well-suited for our cloud-forming numerical experiments \citep{mic08}.

\section{Orders of magnitude}

Order-of-magnitude reasoning suggests the possibility of precipitation close to Martian vapor sources.

Consider liquid water at the surface of Mars: the atmospheric temperature is similar to today. The surface is not entirely frozen because drainage or filling stirs the lake and mechanically disrupts the ice cover, convection mines heat from an underlying hot layer to balance evaporative cooling, subsurface discharge outpaces evaporation, or fumaroles and gas-charged fountains inject vapor and small droplets directly to the atmosphere. The injection rate is approximately \citep{ema94}

\begin{equation}
{Q_h \sim C_D \, \vline \, V_a \, \vline \, (r^*_s - r_b)}
\end{equation}

where $C_D$ $\sim$ 10$^{-3}$ is a surface exchange coefficient (\citet{ema94}, p.484), $V_a$ $\sim$ 10 ms$^{-1}$ is anemometer-level wind speed, $r^*_s$ $\sim$ 0.5 is the near-surface vapor mixing ratio, and $r_b \simeq$ 0 is the background water vapor mixing ratio. To convert this to the vapor mixing ratio in air that enters the buoyant plume we require a length scale (a vertical distance over which the vapor is mixed) and a timescale (during which the vapor is injected). A reasonable length scale is the thickness of the subcloud layer $\Delta Z_{subcloud}$ $\sim$ 2 km. A reasonable time scale is the fetch timescale  $t_{fetch} = D_{lake} \, / \, \vline \, V_a \, \vline  \sim$ 2 hours for a 65km lake (where $D_{lake}$ is lake diameter). This gives a vapor mixing ratio in air that enters the buoyant plume, $r$ $\sim$ 0.01 $\equiv$ 6 Pa.

Convection initiation is made more likely by Mars' low atmospheric temperatures. Air containing 6 Pa vapor will be supersaturated with respect to ice when T $<$ 227K \citep{hardy}. Mars today has an atmospheric surface layer 20K on average colder than its ground surface temperature (Mars Climate Database v4.3, described by \citet{lew99}; \citet{dddmcd}; henceforth Mars Climate Database).  For example, at L$_s$ = 255$^\circ$ (near perihelion) the latitudinal maximum zonal mean daily maximum surface temperature is 304K, but the zonal mean daily maximum atmospheric surface layer temperature at this latitude is only 265K. This offset is due to the low atmospheric column density, because this reduces radiative and mechanical coupling between the atmosphere and surface. In the current climate, diurnal mean equatorial atmospheric temperature is below 227K at all altitudes and all times of the year (Mars Climate Database). Because of these low temperatures, supersaturation of the vapor (and nucleation and growth via deposition) occurs close to the ground. On Earth, the strongest lake storms are associated with rare cold-air outbreaks and large air-lake temperature contrasts \citep{mar10}. On Mars, these conditions occur everyday.

Mars' low atmospheric pressure promotes deep convection (Figure \ref{EQUIVTEMP}). Because of the low volumetric heat capacity, we assume that the plume will accelerate upward until 90\% of the vapor has crystallized. Assuming that no precipitation occurs, with a measured Mars adiabatic lapse rate of $\Gamma_{Mars} \sim$ 1.5 K/km (Mars Climate Database) and the Clapeyron slope for ice at 230K, this occurs 12 km above cloudbase.

During this ascent, the plume will have gained Convectively Available Potential Energy (CAPE):

\begin{equation}
{CAPE \approx  \rho_{air}  g_{Mars}\int^{MC}_{CB} \! \frac{T}{T'} - (1 + \mu) \, \mathrm{d}z }
\end{equation}

where $MC$ is the elevation of almost-complete condensation, $CB$ is cloud base, $T$ is temperature within the plume, $T'$ the environmental temperature, and $\mu$ the ice mixing ratio \citep{rog89}.  This assumes that there is no precipitation of ice out of the parcel. Approximating the ice crystallization as linear from 0 at $CB$ to complete at $MC$, we obtain $T - T'$ = $ r L_v / c_{CO2} $ $\sim$ 30 K so $ T/T'$ $\sim$ 1.15 upon complete crystallization.
$\mu$ $\sim$ 0.01 can then be set aside as negligible. Thus $CAPE$ gained during ascent $\sim$ 3000 J/kg and peak vertical velocity $W_{max}$ = $\sqrt{2 CAPE / \rho_{air}}$ $\sim$ 80 m/s. We have ignored differential pressure gradient acceleration and compensating downward motions \citep{rog89} in obtaining this result. The plume will continue to ascend well above $MC$, but will decelerate as it entrains more ambient air and spreads to form an anvil cloud. The ascent timescale $t_{ascend}$ is 12 km / $0.5 W_{max}$ = 300s. Attention now shifts to the growing ice crystals.

Although Mars air is cold enough at these altitudes to allow homogenous nucleation, we assume heterogenous nucleation occurs on dust. Equatorial dust opacity $\tau_{d}$ is typically 0.01-0.05 during northern summer and 0.1-0.8 during southern summer \citep{liu03}. The present day typical low-latitude effective radius of dust is $\sim$1.6$\mu$m \citep{wol03}. A uniform dust density in the lower 30 km implies a number density of 4 x 10$^4$ m$^{-3}$ (4 x 10$^6$ m$^{-3}$) ice nuclei (considering geometric cross-section only and neglecting self-shadowing) for $\tau$ = 0.01. At 20 km elevation (3 g/m$^3$ air; 0.03 g/m$^3$ H$_2$O), this yields an `seeder' crystal radius $r$ of 58 $\mu$m (or 13 $\mu$m for $\tau$ = 1) -- a minimum, in that it assumes all ice nuclei are consumed. Since crystallization is implicit in the updraft-velocity calculation, we do not assign a separate timescale to nucleation and early growth.

How fast will crystals grow to precipitable size? Assuming that the `seeder' crystals sink relative to the updraft and anvil cloud while scavenging both vapor and smaller droplets, the growth rate is just \citep{rog89}

 \begin{equation}
{\frac{\mathrm{d} R}{\mathrm{d}t} \approx \frac{\bar{E} M}{4 \rho_{i} } \Delta W }
\end{equation}

where $\bar{E}$ $\approx$ 1 is collection efficiency, $M$ $\sim$ 0.03 g/m$^3$ H$_2$O is cloud total water content at 20 km, $\rho_{i}$ = 910 kg/$m^3$ is ice density, and $\Delta W$ = 10 ms$^{-1}$ is a characteristic sink rate relative to the surrounding vapor-laden air. This gives a growth time $t_{growth}$ $\sim$ 2000 s to precipitable size (assumed 200 $\mu$m). In Earth thunderstorms this also requires tens of minutes.

The crystals now begin to fall. At the low temperatures encountered at high altitudes on Mars, we expect the water to form hexagonal cylinders \citep{wal06}. The drag coefficient $C_D$ for cylinders is $\sim$1 over a wide range of Reynolds number $Re$, 10$^2$ - 10$^5$ \citep{tri88}. Therefore we obtain a terminal velocity of \citep{tri88}

\begin{equation}
{ u_{ \infty } =  \sqrt{\frac{2 g_{Mars} \rho_i \pi r}{0.5 \rho_a}} }
\end{equation}

which with $r$ = 200 $\mu$m gives 40 $m s^{-1}$. Using the dynamic viscosity of CO$_2$ at 233K, $\mu_{a}$ = 1.2 x 10$^{-7}$ Pa s, we obtain $Re$ = $\rho_a u_{ \infty } r / \mu_{a} $ $\sim$ 200, sustaining the assumption of 10$^2$ $<$ $Re$ $<$ 10$^5$. Fall time from anvil cloud height is then $t_{fall}$ = $H_{plume}/u_{ \infty }$ $\sim$ 500 seconds.

The total lifetime of the vapor from release to precipitation as snow is $t = t_{ascend} + t_{grow} + t_{fall} \sim$ 3000 s. During the entire process of ascent through the plume, crystal growth, and snow fall, the parcel has been blown sideways by the regional winds. Taking 20 $m/s$ as a representative shear velocity at cloud-forming altitudes of $\sim$ 20 km (Mars Climate Database), we find that snowfall will be roughly $\sim$ 60 km downwind of source. This is small compared to the sizes of many Martian geographic features (craters, canyons, volcanoes). Therefore, provided latent heating powers a strong buoyant plume immediately downwind of the lake that lofts the released vapor to a height at which it will condense, we hypothesize that localized precipitation can occur on Mars.

In order to test this hypothesis, we carry out a set of numerical experiments.

\section{Model setup}
Our numerical experiments used four nested grids with 160 km resolution on the outermost (hemispheric) grid, increasing to 5.9 km on the innermost grid. The lake is centered at 6.5S, 299E (the location of Juventae Chasma; paper 2).

For these simulations we prescribed flat land at 0m on all grids, with uniform albedo = 0.16, uniform thermal inertia = 290 kieffers \citep{mel00}, and uniform roughness $z$ = 0.03m (the value determined for the Mars Pathfinder landing site; \citet{sul00}). We introduced an isothermal lake with albedo = 0.05, and constant roughness $z$ = 0.01m, which is close to the time-averaged roughness value obtained with the sea surface roughness parameterization of Eq. 7.21 of \citet{pie02}. Lake surface temperature is pinned to 278.15K, with saturation vapor pressure according to \citet{hardy}. Constant lake surface temperature is a reasonable approximation if either (1) the lake is deep and well-mixed (e.g. for the 4km-deep lake at Juventae Chasma, cooling rate $\sim$0.01 K day$^{-1}$ for an evaporation rate of 2 mm/hr if the lake is well-mixed), or (2) the lake surface is constantly refreshed by discharge of warm, perhaps gas-charged water from an aquifer \citep{har09,bar10}. In either case, the lake temperature will change more slowly than the (strongly diurnal) atmospheric response. The focus of this first study is to identify the steady-state response of the atmosphere to the lake perturbation, so a time-dependent treatment  of coupled lake thermodynamics is not appropriate. We use the NASA Ames MGCM \citep{hab93} to provide atmospheric boundary conditions. To prevent extremely high water substance mixing ratios and to allow metastable surface liquid water at +2 km above datum (the elevation of the Juventae plateau streams in Paper 2), we double atmospheric pressure relative to today in our simulation by doubling initial and boundary pressures supplied by the GCM. Runs are at L$_s \approx$  270$^\circ$ (southern summer solstice). Runs were for 7 days, except for the \texttt{ref} simulation which was extended to 12 days to test for precipitation variability. Many of these parameters were varied in sensitivity tests: see Table 1 for a list of runs and parameters varied. More details on the model setup are provided in the Appendix.

We carried out a dry run forced by these boundary conditions using flat topography but no lake. In the dry run, surface pressure is 1190-1270 Pa, 40\% more than the saturation vapor pressure at 278.15K. Surface temperatures range from 209-281K. Surface winds are from the NNE, backing to the N during the passage of each afternoon's thermal tide. Wind direction rotates anticlockwise with increasing height until, near cloudtop elevation ($\sim$ 30 km in the reference run), wind direction stabilizes at E to ESE (the subtropical jet of Mars' single Hadley cell). There are no significant day-to-day variations in the wind field. The mean 0-6km shear is 22 m s$^{-1}$.

\section{Analysis of reference simulation}
For our reference simulation, referred to here as \texttt{ref}, we introduce a square lake with sides $\approx$ 65km. This is similar to the diameter of Mojave Crater, a young crater with a fluvially-modified rim \citep{mce07,wil08}). We begin the analysis by describing the time-averaged response.

\subsection{Time-averaged response}
Because the Mars daily average temperature is below freezing, the lake transfers sensible heat to the atmosphere on average. It also injects radiatively-active water vapor into the atmosphere, whose condensation at altitude releases latent heat and produces strongly-scattering water ice aerosols. Time-averaged surface temperatures up to 100 km downwind of the lake are raised by 8K as a result of these effects (Figure \ref{TIMEAVERAGES}a).

This number is small because the  time averaged longwave heating (vapor greenhouse and cloud reradiation) and shortwave cooling (ice scattering) due to the lake nearly cancel. Time-averaged longwave forcing peaks at +106 W/m$^2$ immediately downwind of the lake, which is where vapor column abundance is greatest. Time-averaged shortwave forcing peaks at -83 W/m$^2$ about 20km further SSE, which is where ice column abundance is greatest. Net radiative forcing is $\sim$ +30 W/m$^2$ over the lake, but typically between -3 and +15 W/m$^2$ in the area of greatest ice precipitation (Figure \ref{TIMEAVERAGES}b). Because of the near-uniform northerly wind, we can understand the spatial structure of the net radiative forcing in terms of the timescales needed for vapor to ascend and precipitate out:- net forcing is positive just downwind of the lake, where vapor is still ascending to condensation level; it is modestly positive or slightly negative while vapor is being converted into scatterers; and it becomes positive again when those scatterers have had time (equivalently, distance South) to precipitate out.

A vertical slice through the atmosphere shows three components to the time-averaged atmospheric temperature response. Atmospheric temperatures in a thin boundary layer above the lake are warmed by 30K. The planetary boundary layer is thickened downwind of the lake because of the increased turbulence associated with the lake, so temperatures rise by 4K in the part of the atmosphere that is included in the boundary layer as a result of the presence of the lake. Most importantly, a narrow plume of 5-10K increased temperatures extends 10-15 km above the lake. This thermal plume corresponds to the latent heat released by the lake-induced storm. Atmospheric temperatures never exceed 273K, so liquid water droplets are never stable. Supercooled water droplets are not included in our simulations, and could only form very close to the lake surface.

Strong low-level convergence results from the release of latent heat (Figure \ref{TIMEAVERAGES}d).
Because the lake effect creates its own convergence, this suggests that precipitation should be insensitive to changes in the regional windfield.

The total water column abundance (thin lines in Figure \ref{TIMEAVERAGES}e) shows the narrow extent of the weather system induced by the lake: its core is similar in horizontal extent to a terrestrial thunderstorm. The contours of (ice/total water) fraction (thick lines in Figure \ref{TIMEAVERAGES}e) are extended to the west of the lake because the ice-rich uppermost levels of the cloud are affected by the Easterly subtropical jet. Resublimation within this jet lowers the ice fraction with increasing distance to the west.



\subsection{Spatially-averaged time dependencies}

The area of peak water-ice precipitation is immediately S of the lake. Within a square with sides $\approx$65km immediately S of the lake, the daily temperature cycle is regular, with little variability. Water vapor and cloud blanketing raises nighttime surface temperature by up to 18K relative to the dry run, but the net increase in daytime surface temperature due to the lake is small or negative because of ice-particle scattering (Figure \ref{SPATIALAVERAGES}a). Despite ice-particle scattering, afternoon surface temperature exceeds 273K in the area of peak water-ice precipitation, so snow falling onto bare ground during the afternoon will melt. Snow falling onto ground that is cooled by the increased albedo of snow that has fallen during the night may or may not melt, depending on the grainsize, thickness, and dust content of the nighttime snow layer \citep{clo87}.

To track water and ice mass budgets, we average over a square 400 km on a side which contains the lake but is centered 100km S of the lake in order to enclose the cloud. The time dependence of the water substance mass budget is dominated by a strong afternoon peak in atmospheric water vapor ($\sim$2.5 x the predawn minimum of $\sim$9 Mton) at a time when atmospheric temperatures are highest (Figure \ref{SPATIALAVERAGES}b). The mass of water ice in the atmosphere is independent of time-of-day and averages 10 Mton during sols 5-7. The majority of the water vapor injected into the system precipitates as water ice during the night (see \S4.4). Again, notice that atmospheric temperature does not exceed 273K, so liquid water aerosol is never stable.



\subsection{Structure of the buoyant plume}
Latent heat release drives plume ascent. Figure \ref{PLUME} is a predawn snapshot: water ice mass ratio in the plume core exceeds 1\%. At 17.5 km elevation, more than 95\% of water substance is in the condensed phase and vertical velocities reach 54 m s$^{-1}$. There is now little energy to be gained from further condensation, so the plume slows and broadens. Sublimation, entrainment, and especially precipitation, all lower the ratio of ice to vapor. The plume overshoots its equilibrium level, peaking near 35km.

At a given altitude, plume updraft velocities are strongest just before dawn and weakest in early afternoon. This diurnal cycle in plume behavior corresponds to three related changes in the state of the atmosphere just upwind of the lake. Firstly, the Planetary Boundary Layer (PBL) pinches and swells during the diurnal cycle. The greater depth and intensity of turbulent mixing during the day leads to more entrainment of moist air by the ambient atmosphere, where it does not contribute to the plume. Higher air temperatures disfavor crystallization at low altitudes during the day. Together, these two changes allow more vapor to escape during the day, by advection downwind within the thickened PBL, and not contribute to the plume. Finally, during the day, the excess of lake temperature over land temperature is small, and convergence is weak, so the shear velocity $u_{*}$ is small. Therefore, relatively little vapor mixes from the lake surface boundary layer into the atmosphere. During the night, the greater land-lake temperature contrast is associated with stronger convergence. More vapor is mixed above the surface layer and entrained by the plume. Allowing for dynamic lake surface roughness ($z_0 \sim u_{*}^{2}$) increases the sensitivity of plume velocity to time-of-day, which confirms the effect of surface-layer dynamics on the diurnal cycle.

\subsection{Precipitation}
Precipitation is strongly peaked just downwind of the lake. Figure \ref{PRECIP}a shows that time-averaged snowfall is everywhere $<$ 0.006 $\times$ its peak value at distances $>$ 150 km from that spatial peak. Peak snowfall (Figure \ref{PRECIP}b) shows a similar, but more noisy, pattern, with peak precipitation $>$ 4.5 mm/hr only in a small area beneath the buoyant plume. Precipitation rates increase by a factor of 4 during the night, with a rapid decline during the morning to a lower, stable afternoon rate (Figure \ref{PRECIP}c).
%

Diurnally-averaged precipitation is steady in location over the 12 days of our extended \texttt{ref} simulation. The total precipitation rate increases slightly with time.


\subsection{Precipitation efficiency}
Figure \ref{TERNARYFATE} and Table 2 show the fate of released vapor at the end of sol 6 in \texttt{ref}. The majority of the vapor precipitates $<$200 km from the lake. Of the vapor that reaches distances $>$200 km from the lake, the majority is in the form of vapor at the end of the simulation.

After 6 days of the \texttt{ref} run, the outermost (hemispheric) grid contains $\sim$300 Mt more atmospheric water than the \texttt{dry} run (please refer to Table 1 for details of runs and parameters). This is equivalent to a global surface liquid water layer 2$\mu$m thick, or a lake depth of 7 cm. This is radiatively unimportant on the global scale, and less than the current global average ($\sim$17 pr $\mu$m in the northern hemisphere, $\sim$9.5 pr $\mu$m in the South: \citet{smi02}). The added vapor will presumably precipitate as ice on the winter pole, which is outside the space and time limits of our simulation. If all vapor precipitates in a single season on a polar cap of area 10$^6$ km$^2$, it would form a layer 0.3 m thick. Layers of this thickness could be resolved by the HiRISE camera on a gently-sloping exposure.

Figure \ref{TERNARYFATE} understates the fraction of water that precipitates locally if the conditions that maintain a liquid surface are maintained indefinitely. This is because some of the water that is in the atmosphere at the end of sol 6 will precipitate locally. In an extended run (\texttt{ref (sol 12)} in Figure \ref{TERNARYFATE}), the fraction of water that precipitates locally is increased.

%

\section{Sensitivity tests}

\subsection{Vertical resolution}
We carried out a sensitivity test to determine model vapor release as a function of vertical resolution.
The penalty of increasing vertical resolution scales as $N$log($N$), so runs were for only $\frac{1}{4}$ sol starting at $\sim$ 12 noon Local Solar Time. Vapor release rate is moderately sensitive to model vertical resolution. Increasing surface layer thickness by a factor of 30 decreases vapor release by 35\% (Figure \ref{VERTRES}). Therefore, our modeled afternoon vapor release (and precipitation) rates are probably underestimates. (In an earlier report (\citet{kitlpsc}), we stated the results of a sensitivity test carried out at night, for which the sensitivity has similar magnitude but opposite sign.)

\subsection{Horizontal resolution}
We carried out a simulation, \texttt{hires}, that added an inner grid with 2.0 km horizontal resolution. (\texttt{hires} was initialized from \texttt{ref} output after 2 days). Overall storm structure was similar, although secondary plumes developed in addition to the main plume. Peak vertical velocity increased by 34\%, peak time-averaged vertical velocity by 57 \%, and cloud height by 12\%, relative to \texttt{ref}. However, when averaged to the lower resolution of \texttt{ref}, these differences decrease to 16\%, 26\%, and 12\% (unchanged), respectively (Figure \ref{SIZETEST}). We conclude that our main results are not affected by horizontal resolution. Simulations which use resolutions intermediate between eddy-resolving and the mesoscale are known to suffer from artifacts caused by aliasing of barely-resolved eddies by the grid cell size. Given that we observed 2$\Delta$x noise in \texttt{hires}, we consider our reference run to be more accurate.

\subsection{Size of idealized lake}
Small lakes in our simulations are unable to drive deep convection, and have weaker localized precipitation. We modeled square lakes with areas of $\sim$35 km$^2$ (\texttt{s}), $\sim$300 km$^2$ (\texttt{m}), $\sim$1700 km$^2$ (\texttt{l}), and $\sim$29000 km$^2$ (\texttt{xxl}), in addition to the $\sim$4000 km$^2$ \texttt{ref} simulation.

Deep moist convection was not observed in the \texttt{s} and \texttt{m} simulations. To verify that this was not an artifact of insufficient model resolution, we ran a 2km-resolution nested grid on \texttt{m}. Although peak vertical velocities did increase in the nested grid relative to the default-resolution model, lake-sourced vapor did not reach altitudes much greater than the planetary boundary layer and, as in the default-resolution model, was passively advected downwind. Given our particular choice of atmospheric boundary conditions (present day orbital conditions, low latitude, and southern Summer), lake size $>$ 10$^3$ km$^2$ is required for deep moist convection. If vapor is funnelled to a single central plume of constant radius and entirely condensed, we would expect plume vertical velocity to scale linearly with lake area. Figure \ref{POWERLAW} shows that this expectation is borne out for lakes of size 10$^{2.5}$ km$^2$- 10$^{3.5}$ km$^2$, but overestimates lake storm convective intensity for the largest simulation. This \texttt{xxl} simulation is unusual because it has a much broader area of strong upwelling than the smaller lakes, which explains why its convective velocities are not as high as expected. In the smallest lake modeled, the lake-associated updrafts are so weak that they were difficult to seperate from everyday planetary boundary layer convection.


Because lake-induced convergence (Figure \ref{TIMEAVERAGES}d) efficiently funnels lake-released vapor into one buoyant plume, plume velocity and plume height increase with lake size (Figure \ref{SIZETEST}). Greater plume heights promote ice formation, and the ice fraction of atmospheric water increases with lake size (colors in Figure \ref{TERNARYFATE}). This is a self-sustaining feedback, because ice formation provides energy for plume ascent, which in turn creates low pressure above the lake and drives convergence.

Larger lakes inject proportionately less water into the global atmosphere. More than $\frac{1}{3}$ of the water released in the \texttt{s},\texttt{m} and \texttt{l} simulations is in the atmosphere at the end of the simulation. $<$14\% is in the form of ice. In the \texttt{xxl} simulation, the fraction of released water that is precipitated locally is $>$90\%, and most of the atmospheric water is ice aerosol that would probably precipitate locally if the lake surface froze over. The \texttt{line} source has a similar lake area to the square \texttt{xxl} run, and behaves similarly in having a small atmospheric water fraction and a large ice aerosol fraction. This suggests that this trend of increased localized precipitation fraction with increased lake area does not depend on lake geometry.

\subsection{Surface roughness parameterization}
Our default run assumes fixed lake surface roughness, but water surface roughness increases with wind speed. A more realistic approximation is that of \citet{gar92} as given by Eq. 7.21 of \citet{pie02}:

\begin{equation}
z_0 = 1 \times 10^{-4} + (0.01625 u^{2}_{\*} / g)
\end{equation}

where we have added the first term to maintain numerical stability.
This remains an adequate fit to the much larger datasets now available \citep{toga_coare_comparison}.
The time-average lake roughness with this parameterization is 0.0126 m, and is a factor of three higher during the night when near-surface winds are stronger.

Our time-averaged results are insensitive to this more accurate lake surface roughness parameterization (Table 2; Figure \ref{TERNARYFATE}), although the very strong plumes seen during the night (when vapor injection rate is highest) increase both plume height and updraft velocity (Figure \ref{SIZETEST}). 

\subsection{Line source}
Our line source is intended to sketch an outflow channel during a channel-forming flood. It is a N-S oriented, straight trough of depth 1.5 km, floor width 30 km, wall slope 0.13, and length 880 km. The floor is flooded. Of the resulting snow, 23\% falls back into the lake and a further 70\% falls within 100 km of the edge of the trough. The diurnal cycle consists of a strong, steady, spatially continuous line storm on the W edge of the trough during the night, and a clumpy, broken line of weak updrafts some distance E of the trough during the day. Time-averaged vertical velocity and precipitation fields do not show this clumpiness: intead, there is a trend of monotonically increasing vertical velocity and precipitation rates toward the north, because the background wind field advects the upper parts of the cloud toward the north.

We do not consider the influence of the flowing water on surface roughness, nor the drag of the flowing water on the atmosphere.

\subsection{Season}

In southern winter (also aphelion season) the sign of the Hadley circulation reverses. In a run at Ls = 90$^\circ$, the GCM boundary conditions produced ESE-directed time-averaged winds at altitudes below $\sim$ 15 km, and SW-directed time-averaged winds at higher altitudes. The highest snowfall rates were just ESE of the lake, reflecting steering of the plume by regional winds. Total localized precipitation was the same as in \texttt{ref} to within 7\%.

\subsection{Latitude}
An cyclonic circulation does not develop in our \texttt{ref} run. To determine if Coriolis effects can produce an cyclone at higher latitudes, we ran a test at 45$^\circ$S latitude. As expected given the small size of the lake, the lake-driven circulation is too weak to restructure the background wind field and a lake-induced standing cyclone does not form.

The total localized precipitation in the midlatitude run is only $\frac{1}{3}$ of the total localized precipitation in the equatorial runs. Since vapor release is similar, vapor is being converted to localized precipitation with a smaller efficiency. Peak time-averaged column ice fractions are $<$ 40\% in the midlatitude run, compared to $>$ 70\% in the reference run (Figure \ref{TIMEAVERAGES}e). This is because both runs are close to southern summer solstice. The southern midlatitude site is in sunlight for $2/3$ of each day, and has air temperatures higher by 20K on average than the equatorial site. Because of the lower supersaturations, water in the midlatitude plume must be lifted 5 km higher to obtain a given ice fraction than in the equatorial plume. Therefore, more vapor escapes to the regional atmosphere.



\subsection{Faint young sun}
Many channel networks on Mars formed when the Sun was fainter. The solar luminosity $L$ at the Hesperian-Amazonian boundary was 0.78-0.85 $\times$ present-day solar luminosity $L_{NOW}$, and $L$ at the Noachian-Hesperian boundary was 0.75-0.77 $\times$ $L_{NOW}$ \citep{bah01,harrecent}. The uncertainty is due to differences among the models that map crater density onto absolute age: the solar evolution model has much smaller error bars. The Juventae plateau channel networks (Paper 2) could be as old as Hesperian \citep{led10}, so we ran a test at 0.75 $L_{NOW}$. As well as being less luminous, the young Sun was also 1.5 \% cooler \citep{bah01}. We ignore the resulting small shift in the solar spectrum and simply reduce the flux at all wavelengths by 25 \%.

Colder air temperatures under the faint young sun lower the cloud base by $\sim$ 5 km. Peak updraft velocities are $>$50 m.s. in both simulations, but occur at lower elevations in the faint young sun simulation. Because supersaturations are higher at all colder altitudes, ice growth is favored and the fraction of atmospheric water that is ice increases from 32 \% in \texttt{ref} to 47 \% in \texttt{lo\_sun}.

We conclude that localized precipitation is favored by reduced solar luminosity. However, this is not true for runoff generation and channel formation (\S 6).


\subsection{Paleoatmospheric pressure}

Figure 1 suggests that localized precipitation will not occur if atmospheric pressure is greatly increased. To test this, we increased pressure to 10x Present Atmospheric Level (PAL) on all the mesoscale grids (run \texttt{hipress}). Deep moist convection is suppressed, uplift velocities are reduced, and comparatively little vapor reaches cloud-forming altitudes. No lake-sourced ice is found above 10km, and little ice aerosol forms (1.6 \% of atmospheric water mass on the inmost grid, versus 32 \% in \texttt{ref}).

Precipitation is reduced by 20 \%, and preliminary runs at 20x PAL lead us to expect that further increases in pressure will greatly reduce localized precipitation. However, the mismatch between the forces driving winds on the GCM (at 1x PAL) and and the forces driving winds on the mesoscale grids at 10x PAL is already severe. Higher-pressure simulations will benefit from higher-pressure GCM runs.

\section{A probabilistic model of snowpack melting on ancient Mars}

For a lake lifetime of 1 Earth year (similar to expected chaos outflow durations: \citet{and07,har09}), and assuming that seasonal effects on the storm are minor, $\sim$2 m of snow is predicted to accumulate downwind of the lake by our reference simulation, assuming a snowpack density of 350 kg/m$^3$. Will it melt?

We expect the lifetime of snowpack against sublimation and wind erosion to be $\geq$ $O(1)$ year, so that the annual-maximum temperature is that which is relevant for snow melting. For predicted snowpack depths $O$(1)m, this residence time is supported by GCM simulations (e.g., Table 1 in \citet{mad09}). A simple energy argument gives the same result --

\begin{equation}{t_{sublime} \sim \frac{d_{snow} \rho_{snow} L_{s}}{F_{Sun} \eta f_d} }\end{equation}

which with snow depth $d_{snow}$ = 1 m, snow density $\rho_{snow}$ = 350 kg/m$^3$, solar flux $F_{Sun}$ = 300 W/m$^2$, fraction of solar energy used for sublimation $\eta$ = 0.1, day fraction $f_d$ = 0.5, gives $t_{sublime}$ = 1.0 Mars year. With surface pressures on the plateau of $\approx$1000 Pa, maximum modelled surface shear stress is not sufficient to initiate motion of snow that is resting on the ground \citep{windgeo}, so snow cannot be blown away.

Atmospheric temperatures in our model are too low for rain to be stable, so precipitation falls as snow.
This is true even for runs (not shown) in which initial surface temperature is that of impact ejecta: for a low-pressure, radiatively thin atmosphere similar to today, energy transfer from the surface to the atmosphere is slow and inefficient. If localized precipitation in a thin atmosphere similar to today can account for channel formation, it must be shown that surface liquid water runoff can occur on $<$5$^\circ$ slopes in a thin atmosphere similar to today. Midlatitude channels at Acheron $\sim$0.08 Ga and Lyot $<1.5$ Ga formed on $<$5$^\circ$ slopes sufficiently recently that it's unlikely that the atmosphere when they formed was much thicker than it is today \citep{dic09,fas10}. This offers empirical support for the snowpack runoff hypothesis. However, Amazonian midlatitude channels are not common, and formed in unusual circumstances (such as nearby steep slopes, and glaciers) \citep{fas10} that may not hold for low-latitude, localized channels. In addition, some of the observed low-latitude channels (e.g, those described in Paper 2) formed $>$ 3.0 Ga when solar luminosity was $\le$80\% that of today's Sun \citep{bah01}. A model of the likelihood of snowpack melting is needed.

In general melting probability will depend on orbital elements including obliquity and precession, latitude, age (via solar luminosity), material properties, and the lifetime of snowpack against sublimation losses to planetary cold traps. For age $\ge$20 Mya, deterministic chaos makes orbital elements unreliable: as our interest is in ancient Mars our model is therefore probabilistic. We model temperatures at the equator - 64\% of the sedimentary rocks on Mars are within 10$^\circ$ of the equator (\citet{mal10}, \texttt{figure16.txt} in their online supporting data), and our modeling target in Paper 2 is at 5$^\circ$S.

\vspace{0.05in}

\noindent
{\it a. A a simple energy balance.} The energy balance of a snow-bearing surface on Mars is \citep{dun10}

\begin{equation}
{\frac{\partial U}{\partial t} = (1 - \alpha)SW\hspace{-0.05in}\downarrow + \epsilon LW\hspace{-0.05in}\downarrow - \epsilon \sigma T^4 - SH - \- C_{net} - L_{sub} \frac{\partial m_w}{\partial t}}
\end{equation}

where the left-hand side is gain of energy by the surface layer, $\alpha$ is surface albedo, $SW\hspace{-0.05in}\downarrow$ is sunlight, $\epsilon$ $\sim$ 0.98 for ice is thermal emissivity, $LW\hspace{-0.05in}\downarrow$ is the greenhouse effect, $\sigma$ is the Stefan-Boltzmann constant, $T$ is surface temperature, $SH$ is net sensible heat flux from the surface to the atmosphere, $C_{net}$ is conductive heat flux from the surface into the ground, $L_{sub}$ is the latent heat of sublimation of water, and $\frac{\partial m_w}{\partial t}$ is the sublimation rate.

The minimum sublimation rate (in the absence of wind) is given by \citep{hec02,dun10}

\begin{equation}
{\frac{\partial m_w}{\partial t} = 0.14 \Delta \eta \rho_{ave} D \left( \left( \frac{\Delta \rho_{air}}{\rho_{air}} \right) \left( \frac{g_{Mars}}{\nu^2_{air}} \right) \left( \frac{\nu_{air}}{D_{air}} \right) \right)^{1/3} }
\end{equation}

where $\Delta \eta$ is the difference between atmospheric and surface gas water mass fractions, $\rho_{ave}$ is the atmospheric density, $D_{air}$ is mass diffusivity, $\Delta \rho_{air} / \rho_{air}$ is the normalized density difference between the moist near-surface atmosphere and ambient atmosphere, and $\nu_{air}$ is kinematic viscosity. We parameterize the temperature-dependency of $\nu_{air}$ and $D_{air}$ following \citet{dun10}.

Melting occurs if $\frac{\partial U}{\partial t} \ge$ 0 at $T$ = 273.15K. We set up optimistic conditions for melting, and ask: with what probability do orbital conditions permit melting?

 Evaporative cooling is handled following \citep{dun10}: we obtain 165W assuming 70\% humidity, atmospheric pressure 1220 Pa, and zero wind. The more sophisticated surface-layer theory of \citet{clo90} gives an evaporative cooling rate $>$2 times less than in \citet{dun10} (and $>$4 times less than in \citet{ing70}) for roughnesses $z_0$ $<$0.3 mm that are appropriate for polar snow \citep{bro06}. Therefore, our parameterization may overestimate evaporative cooling by free convection. We assume still conditions, so that cooling by wind-driven turbulent fluxes (forced convection) is negligible: this assumption is favorable for melting. We also neglect the turbulent flux of sensible heat $SH$, which is a factor of $>$10 smaller than evaporative cooling for these conditions \citep{hec02}. The noontime conductive heat flow $C_{net}$ can be parameterized assuming the temperature falls to the diurnal average temperature at a depth equal to twice the diurnal skin depth. Assuming a diurnal-average temperature during the hottest season of 220K, this gives

\begin{equation}
{C_{net} = k \frac{\partial T}{\partial z} = k \frac{273.15 - 220}{2 \sqrt{\kappa P/\pi} }}
\end{equation}

For thermal conductivity $k$ = 0.125 W/m/K for snow \citep{car03}, sol length $P$ = 88775.204 s, and thermal diffusivity $\kappa$ = 2.04 x 10$^{-7}$ m$^2$s$^{-1}$ appropriate for low-density snow, we obtain $C_{net}$ = 40.3W/$m^2$.

The greenhouse effect $LW\hspace{-0.05in}\downarrow$ at the time of peak surface temperature is assumed to be 55W/m$^2$ (Mars Climate Database). We do not consider the additional greenhouse forcing from the lake storm, nor from a possible past stronger greenhouse effect.

For dusty snowpack with albedo 0.28, the resulting minimum on $SW$ is 640W.

We calculate the maximum equatorial luminosity for all ages, seasons, and orbital conditions. The age-dependent solar luminosity is taken from the standard solar model of \citet{bah01}. We then weight the annual-maximum luminosity results by the age-dependent probability densities for obliquity and eccentricity from \citet{las04}, assuming that obliquity and eccentricity are not strongly correlated. (In the 0.25 Gyr solutions provided by J. Laskar at \texttt{http://www.imcce.fr/ Equipes/ASD/ insola/mars/mars.html}, obliquity and eccentricity are not strongly correlated.) 640W is exceeded with 21\% probability at 1 Gya, but with only 0.5\% probability 3 Gya (Figure \ref{EXCEEDANCE}).

\vspace{0.05in}

{\it b. A 1D column model.} Next we calculate temperatures within snowpack for the full range of obliquities ($\phi$, 0$^\circ$ $\rightarrow$ 80$^\circ$), eccentricity ($e$, 0.0 $\rightarrow$ 0.175), longitudes of perihelion ($L_p$) and solar longitudes that are sampled by Mars \citep{las04}. We use a 1D thermal model to calculate temperature $T$ within snowpack at a range of seasons, ignoring the greenhouse effect $LW\hspace{-0.05in}\downarrow$. $T$ is solved for by matrix inversion for the linear (conduction) part and iteratively for the nonlinear contribution of radiation and evaporative cooling in the topmost layer. The timestep is 12 seconds. Evaporative cooling is handled following \citep{dun10} with optimistic assumptions as in the simple energy balance. Each run is initialized with surface temperature $T_{surf}$ at instantaneous thermal equilibrium with incoming sunlight, decaying with an e-folding depth equal to the diurnal skin depth $d = \sqrt{(k \rho) / (P c_p \pi)}$ to an energy-weighted time-averaged equilibrium temperature at depth. (Here, $k$ is thermal conductivity, $\rho$ density, $P$ the length of 1 sol in seconds, and $c_p$ specific heat capacity of the subsurface.) We then integrate forwards in time for multiple sols, but with seasonal forcing held constant, until sol peak $T_{surf}$ has converged. Models that neglect the greenhouse effect provide a good approximation to Mars' observed surface temperature.

Atmospheric contributions $LW\hspace{-0.1in}\downarrow$ are parameterized as a time-independent greenhouse temperature increase, $\Delta T_{at}$.
The purpose of adding the atmospheric contribution in postprocessing is to allow different models of the past atmosphere to be compared to each other without rerunning the ensemble of $\sim$10$^5$ column models.
Typical present-day values are $\Delta T_{at}$ = 5-10K \citep{rea04}. This averages over strong nighttime warming, which is irrelevant for melting, and weak afternoon warming, which is critical to melting. Setting $\Delta T_{at}$ to the time average of present-day greenhouse warming will therefore lead to an overestimate of melting. Instead, we want to know the value $\Delta T_{at}$ at the peak temperatures relevant to melting. We make the approximation

\begin{equation}{\Delta T_{at} \simeq T_{surf,max} - \sqrt[4]{\sigma T_{surf,max}^4 - \frac{LW\downarrow \epsilon}{\sigma} }} \end{equation}

The Mars Climate Database Mars Year 24 simulation shows $\Delta T_{at}$ 5-7K at low latitudes. This excludes the dust storm season, for which the neglect of atmospheric scattering leads to an overestimate of the net atmospheric contribution. We therefore adopt 6K as our greenhouse forcing. Additional greenhouse forcing due to doubled CO$_2$ is small and safely neglected \citep{wordsworth}.

Material properties for snowpack are taken from \citet{car03}: we assume a snowpack density of 350 kg/m$^3$, thermal conductivity of 0.125 W/m/K, and specific heat capacity $c_p$ of 1751 J kg$^{-1}$, and neglect the temperature dependence of $k$ and $c_p$. We consider dirty H$_2$O snow albedos between 0.28 and 0.4. The smaller value is the present-day albedo of Mars' dust continents \citep{mel00}, and the larger value is the upper end of the envelope that best fits the observed seasonal dependence of ephemeral equatorial snow on present-day Mars \citep{vin10}. (Paper 2 provides a more detailed discussion of snowpack albedo).

Maximum-temperature results are interpolated onto a finer mesh in orbital parameters ($\phi,e,L_p$). We again weight the results using the probability distributions of \citet{las04}. Results for albedo = 0.28 and latitude = 0$^\circ$ are shown in Figure \ref{CDFS}. Temperatures greater than 273.15K are not realistic, because snowpack temperature will be buffered by the latent heat of melting. The agreement between the melting probabilities obtained from the 1D column model and the melting probabilities obtained from the simple energy balance argument is excellent (Figures \ref{EXCEEDANCE} and \ref{CDFS}). Even with the weak, present-day greenhouse effect, temperatures high enough for melting occur during 4\% of the years (0.1\% for albedo 0.35). With an additional 6K of greenhouse warming, melting occurs during 56\% of years (12\% for albedo 0.35).

\vspace{0.05in}

{\it Discussion.} The exact melting probabilities are unlikely to be correct, but we draw three lessons from these simple models: (1) The probability of low-latitude melting is certainly higher for instantaneous emplacement of snow than for gradual (e.g, orbital equilibrium) emplacement of snow, and probably much higher. That is because gradual emplacement of annually-persistent snow at the equator is only possible at high obliquity ($\sim$50\% of the time), while instantaneous emplacement can occur at any time. In addition, gradually-emplaced snow will accumulate preferentially in areas where it is most stable: for example, shadowed areas, adverse steep slopes, and especially chasma and crater interiors where shadowing reduces peak temperature. Craters at 10$^{\circ}$S in Sinus Sabeus appear to contain mantled, atmospherically-emplaced ice deposits, providing direct evidence that equatorial ice on Mars does preferentially accumulate in crater-interior cold traps \citep{she10}. The most compelling evidence for low-latitude glaciation on Mars is on the flanks of the Tharsis Montes, which have the lowest atmospheric pressures on the planet \citep{for06}. Liquid water exposed on the Tharsis Montes today would boil internally, leading to very rapid evaporative losses. Taken together, these observations suggest that localized precipitation is much more likely to melt than orbital-equilibrium precipitation.

Today's orbital conditions would lead to melting of snowpack on the equator (diamond in Figure \ref{EXCEEDANCE}). However, on today's Mars, snow is almost entirely restricted to high latitudes. Snow and ice on Mars do not melt in general because the area of transient liquid water stability changes on slow, orbital timescales $O(10^{4-5})$ yr. This allows time for ice to be removed (vertically or latitudinally) from the advancing zone of increased saturation vapor pressures, towards cold traps where melting cannot occur. Therefore the areas of surface ice and of transient liquid water stability rarely intersect. Theory, experiments, and the lack of evidence for transient liquid water at the Phoenix landing site confirm this \citep{schorg,huds,phoenix}. Steep ($>$20$^\circ$) gullied slopes within mid- and high-latitude craters are an exception \citep{gullies}, assuming that the gullies formed from liquid water flows. In contrast, localized precipitation delivers snow on timescales $O(10^{-1})$ yr to a location uncorrelated with transient liquid water stability, so a wider range of orbital conditions will then allow melting.

In the case of impact-induced precipitation, snow falling on hot ejecta will melt regardless of orbital conditions.

(2) Because of the wide range of possible orbital states, the probability of melting is not large, but neither is it zero. The distribution of melting events ($T_{surf}$ $\ge$273K) with orbital parameters is shown in Figure \ref{MONTECARLO}. Relative to the pdf of obliquity, melting events are more probable at low obliquity than high; they are much more probable when perihelion occurs at equinox than when perihelion occurs at solstice; and they occur exclusively at moderate-to-high eccentricity. These results may be understood as follows. Moderate eccentricity is needed to offset the reduced solar luminosity. For example, at 3.0 Ga and e = 0.11, perihelion insolation is eqiuivalent to insolation on a circular orbit today. Perihelion near equinox is needed to align the season when the noontime sun is highest at the equator with the peak solar flux. Increasing obliquity narrows the seasonal window during which the sun rises high enough in the equatorial sky for melting to occur. Therefore, the probability of aligning with perihelion is reduced and so the chance of a melting event at high obliquity is lower.

This distribution is almost exactly a ranking in peak solar luminosity, and is not sensitive to our model assumptions.

(3) Using slightly less optimistic assumptions in the energy balance approach (humidity = 0.25 and albedo 0.35), the minimum on $SW$ is 882W. This is unachievable at Mars. For comparison, the sunlight striking the top of the atmosphere above North Greenland, 80N, at summer solstice during the melt season is 730W. The strength of the evaporative cooling term at low pressures strongly suppresses snowpack melting. Evaporative cooling may not be relevant for understanding the melting of small quantities of snow that is in contact with non-volatile soil, and small quantities may be all that is necessary to redistribute mobile elements and to form crusts in the global soil \citep{amu08,arv10}. However, this cannot account for runoff generation and channel formation. Adding CO$_2$ pressure quickly damps evaporative cooling, but also increases the magnitude of sensible heat transfer to the atmosphere. There is a minimum in cooling at intermediate pressures \citep{hec02}, but this may suppress localized precipitation.

For Early Mars, this dilemma can probably only be resolved by an early orbital state with a smaller semimajor axis or smaller eccentricity than the current one; a multibar CO$_2$/H$_2$O atmosphere; or adding non-CO2 greenhouse gases. In summary, our models do suggest a requirement for a different global climate state if localized precipitation is to lead to substantial runoff on gentle slopes on Early Mars, but this is only necessary to bring temperatures back up to the level of contemporary Mars: melting probabilities with present-day solar luminosity are large (Figure \ref{ALBEDOLUMINOSITY}). To counteract the effect of the Faint Young Sun on mean Mars surface temperature at 3.5 Gya, the required greenhouse warming is approximately

\begin{equation}
{\Delta T_{fys} = (1 - 0.77^{0.25}) T_{Mars} \approx 13K}
\end{equation}

 (with $T_{Mars}$ $\sim$ 210K; the luminosity at 3.5 Ga is from \citet{bah01}). This is a relatively modest requirement. It does not hold for impact-induced precipitation, because snow falling on hot ejecta will melt provided that atmospheric pressure is above the triple point.

%
%

%

%
%
%
%

\section{Prospectus and discussion}

Our model shows lake storm ice forming at terrestrial cirrus cloud temperatures $>$200K, which because of the higher water vapor mixing ratio is warmer than modern Mars cloud forming temperatures $\le$185K. Recent laboratory experiments \citep{ira10} show that ice nucleation at T $\le$ 185K requires greater supersaturations than previously thought, while confirming that the MRAMS-CARMA parameterisation of the critical supersaturation at the higher ice nucleation temperatures observed in our simulation is adequate.

 We have focussed on ephemeral open-water lakes in this study, but this is not the only way of getting water into the atmosphere. Fumaroles and rootless cones inject vapor to the atmosphere, but their efficiency has not been as well studied as that of a wind-stirred lake surface \citep{toga_coare_comparison}. Gas-charged fountains may inject vapor during the early stages of outflow channel formation \citep{bar10}. The mass flux we obtain in even the smallest simulation greatly exceed the greatest spring discharges on Earth.
Leads in lake ice may allow deep convection to continue after the majority of the lake surface has frozen over.

\subsection{Geomorphic paleobarometer?}
Cold lakes at 1 bar on Earth do not produce vigorous storms, with convection and precipitation in a dry background atmosphere (Figure \ref{GREENLANDPOOL}); cold lakes at 12 mbar on Mars do, at least in our model. Therefore there is a threshold atmospheric pressure, between 12 mbar and 1 bar, below which cold lakes on Mars can induce convective storms and localized precipitation (orange line in Figure \ref{EQUIVTEMP}). We speculate this pressure is $\sim$10$^2$ mbar, because this is the pressure at which a 273K lake on Mars produces the same buoyancy flux as 299K on Earth. 299K is the lower limit of the tropical sea surface temperatures that on today's Earth are associated with deep convection \citep{ema94}. A preliminary run at 20x present-day pressure ($\sim$ 120 mbar) showed significant suppression of the buoyant plume, with limited localized precipitation. If a past landform on Mars could be conclusively shown to result from localized, lake-induced precipitation, that would suggest a low atmospheric pressure ($\leq$ 10$^2$ mbar) at the time it formed. We investigate one set of candidate landforms in Paper 2, but because of geological ambiguities it does not reach the `conclusive' standard.

If the vapor injection mechanism is a fumarole or gas-charged fountain rather than a lake surface, then the ability to set a paleopressure constraint goes away. The potentially much higher injection rate can overcome the dilution by the thicker atmosphere.

\subsection{Applications to the pre-modern sedimentary record}
One motivation for our localized-precipitation work is to understand Late Hesperian and Amazonian fluvial features which are widespread, although rare, on the Martian surface (Figure \ref{MOTIVATION}; Paper 2). They postdate the global decline in valley formation, aqueous mineralisation and erosion rates near the Noachian-Hesperian boundary. Therefore, localized processes are {\it a priori} a reasonable explanation for these fluvial features. Unusual microclimates (such as nearby steep slopes) may explain some of the features \citep{fas10}, but those on the Valles Marineris plateau and at Mojave Crater suggest localized precipitation (Figure \ref{MOTIVATION}; \citet{man08}; Paper 2).


Impact-induced precipitation has been proposed to explain fluvial fans at Mojave Crater, but not yet modeled \citep{wil08}. Mojave is a fresh crater whose inner terraces are dissected by channelized fans that drain hillslopes. Several other Amazonian craters show Mojave-like fans, but they are more degraded (e.g., PSP\_007447\_1995 in Tartarus and PSP\_008669\_1905 in Isidis). Mojave has been described as a `Rosetta Stone' for decoding impact-associated fluvial features on Mars, because it is relatively young and records with high fidelity processes that are degraded in the ancient record \citep{seg08,ste08,too10}.

\subsection{Can localized precipitation explain Noachian erosion?}

Early Mars' geomorphic record strongly suggests that long-lived and regional/global precipitation contributed to Noachian erosion, but it is also consistent with a role for localized precipitation. The well-dated Late Noachian / Early Hesperian \citep{fas08date,hyn10} peak in valley formation on Mars includes some integrated drainage networks extending over 1000s of km, that appear to have formed over a long period of time relative to crater formation (e.g., \citet{bar09}). Because localized precipitation on a cold planet is almost always short-lived precipitation (\S 1), these findings are not consistent with the model developed in this paper. In addition to this timescale problem, the rapid discharges and short formation timescales required to form some valley networks and deltas are at best marginally compatible with snowmelt \citep{kra08,kle10}, and suggest ejecta dewatering or rain-phase precipitation. (Atmospheric liquid water aerosol is currently not included in our model.)

On the other hand, the majority of drainage systems on Early Mars are poorly integrated and localized, including most infilled crater floors and eroded crater rims. Inspired by observations at Mojave Crater \citep{wil08}, and supported by the results of the idealized model presented here, we suggest that localized precipitation is relevant to understanding these systems. Mars' deuterium-hydrogen ratio indicates that Mars has lost water over time \citep{jak91}. Therefore, crater lakes resulting from impacts into icy targets, wet-based ice sheets and lava-ice interactions would all have been more common on ancient Mars, and could serve as localized vapor sources. An Early Mars GCM with water sources at the valley-associated crater lakes shows precipitation only near the lakes and at planetary cold traps, consistent with our mesoscale results \citep{sot10}.

Any hypothesis for Early Mars erosion should explain the transition from significant crater infilling and crater rim erosion prior to the Noachian-Hesperian boundary, to minimal erosion in the Mid-Late Hesperian and Amazonian (as documented by, for example, \citet{gol06,for04,boy07,how05,irw05,moo05}). In a localized precipitation scenario, this would correspond to a change in planetary target properties: an ice-rich crust before the erosional transition, and an ice-poor crust afterwards.

For impacts into a planetary crust containing buried ice layers, models suggest that fast-moving ice-rich debris flows cause crater infilling in the minutes after impact \citep{ste08}. Therefore, we focus on relatively late-stage, relatively prolonged (hours to millenia) erosion of the inner slope of the crater rim.

Energy balance suggests that it is physically possible for localized precipitation to contribute to increased Noachian crater rim erosion. The energy source for the system is the heat of shocked target material. If the vapor source close to the crater center (fumarole, geyser or lake) can mine heat from deep within the shocked material, a sustained vapor source for years is possible, but the interval over which snow melts on contact with fresh ejecta on the rim is limited by conductive cooling of the surface boundary layer. However, if snowmelt-fuelled erosion removes the chilled boundary layer, hot material will again be exposed and melting can resume. Therefore, positive feedback between erosion and melting can sustain snowmelt-driven erosional activity on the crater rim. Snow can mine heat from deeper within the ejecta blanket provided that

\begin{equation}{(T_{b} - 273.15) c_{b} > \frac{Wt}{Rk} \left( L_{melt} + (273.15 - T_{sn})c_{sn} \right) }\end{equation}

where $Wt/Rk$ is the water:rock ratio required for erosion, $L_{melt}$ is the latent heat of melting, $T_{sn}$ is snow temperature, and $c_{sn}$ is snow heat capacity.

Ejecta blanket thickness is variable but lunar scalings suggest craters $>$ 160 km diameter have ejecta $>$ 0.6 km thick at their rim \citep{mcg73}. Therefore, if ejecta temperature exceeds 900K, snow is continuously supplied at 1.25 mm/hr as found in our simulations, and the majority of eroded material is removed by debris flows with a $W/R$ of 1.5, condition (14) is satisfied, and $>$ 0.6 km material can be eroded from the inner slope of the crater rim by localized precipitation within a decade.

This is directly relevant to three of the final four Mars Science Laboratory candidate landing sites. Gale (155 km diameter), Holden (155 km diameter), and Eberswalde (65 km diameter; immediately NE of Holden), because all three have deeply dissected rims showing evidence for aqueous transport of material from the rim to the crater floor \citep{and10,moo05,ric10}.
%

\section{Conclusions}

We conclude from this study that:-

(1) A low-pressure lake effect creates deep convection, rapid updrafts, and intense precipitation above cold liquid water surfaces on Mars. On Earth, because of much higher atmospheric pressure, this requires tropical temperatures.

(2) We use MRAMS to simulate lake storms on Mars. The modeled storms have updraft velocities and plume heights which exceed the intensity of the strongest recorded thunderstorms on Earth.

(3) The fraction of vapor that is trapped near the vapor source as localized precipitation increases with lake size.

(4) The localized greenhouse effect of the released water vapor is too weak to cause melting of the snow.

(5) Melting of equatorial, rapidly-emplaced localized snow with the albedo of dust is possible for a subset of orbital conditions, even with the present-day weak greenhouse effect.

(6) Assuming that transient lakes on Mars are uncorrelated with orbital forcing, melting of rapidly-emplaced localized precipitation is more likely than melting of precipitation that has been emplaced in equilibrium with orbital conditions.

(7) Taken together, localized storms, rapidly-emplaced localized precipitation, and favorable orbital conditions provide an alternative working hypothesis for (at least part of) the erosion and channel formation observed on pre-modern Mars.


%
%
\appendix \section{Methods} 
The fully compressible dynamical core of MRAMS is derived from the terrestrial RAMS code. A cloud microphysical scheme derived from the Community Aerosol and Radiation Model for Atmospheres (CARMA) was recently added to MRAMS by T.I. Michaels. This cloud microphysics scheme has been used to successfully reproduce spacecraft observations of low-latitude clouds downwind of the Tharsis Montes and Olympus Mons \citep{mic06}.  A recent description of MRAMS capabilities is \citet{mic08conf}.

We made minimal modifications to the MRAMS v.2.5-01 code to allow for surface liquid water.
Liquid water microphysics is not included.
Water vapor thermodynamics are included in the energy equation, but water vapor is not included in the mass and momentum equations -- that is, we ignore pressure and virtual temperature effects. We do not permit dynamic dust lifting at the mesoscale.

Model vertical layer thicknesses varied from 2.3 km at altitude to 30 m near the ground. We used four grids with the outermost being hemispheric and a horizontal resolution of $\sim$5.9 km on the inmost grid. Output was sampled every 1/24 sol ($\approx$ 3699 s), or `Mars-hour.'  We assume that this frequency, limited by available disk space, is enough to capture model behavior -- for example, we refer to the warmest of 24 samples during a sol as the `day's maximum temperature.' The timestep varied between runs but was never more than 3.75s on the inmost grid.

The season is southern summer (Ls $\sim$ 270$^\circ$) for the runs, and boundary conditions are from the NASA Ames MGCM, version 2.1. We use present-day orbital conditions.

Our spinup procedure included 24 Mars-hours with no vapor release from the lake, 3 Mars-hours with vapor release but cloud microphysics off, and the remainder of the run with aerosol microphysics on. We observed that obvious spin-up transients had died away by the end of day 3. Snow albedo feedback was not considered, but would tend to increase the intensity of convergence and convective intensity by increasing the temperature gradient between land and lake.

%
%
%
%

%
%
%
%

\begin{acknowledgments}
We thank Teresa Segura, Sarah Stewart, and Max Rudolph for useful suggestions, and Keith Harrison and Bob Grimm for commenting on the manuscript.  This work made use of GRaDs scripts by Bin Guan and Chihiro Kodama. We made use of the LMD/Oxford Mars Climate Database v4.3. We acknowledge support from Teragrid allocation TG-EAR100023, NASA Science Mission Directorate grants NNX08AN13G and NNX09AN18G, and NASA grants to SwRI which funded cloud microphysics capabilities.
\end{acknowledgments}

%
%
%
%
%
%
%
%
%
%




%
%

\end{article}







%
%
\newpage
\pagebreak

 \begin{table}[ht]
\caption{List of runs with parameters.}
\centering
{\small
\begin{tabular}{l l c l c c c}
\hline\hline
Run & Full name & Lake size (pixels) & Description & Deep moist convection?\\ [0.5ex]
\hline
\texttt{dry}  & lake\_0.005\_Mar\_7\_2010 & -- & Control case without lake & -- \\

\texttt{ref} & lake\_0.000\_Jan\_10\_2010 & 11x11 & Reference simulation (13 days)& {\bf \textcolor{green}{ \( \mathbf{ \surd } \) } }
\\
\texttt{line}  & lake\_0.001\_Jan\_10\_2010 & 5x149 & Line source in valley  & {\bf \textcolor{green}{ \( \mathbf{ \surd } \) } } \\
\texttt{s}  & lake\_0.018\_Mar\_30\_2010 & 1x1 & Size sensitivity & {\bf \textcolor{red}{ \( \mathbf{ \times } \) }}\\
\texttt{m}  & lake\_0.019\_Mar\_30\_2010 & 3x3 & Size sensitivity & {\bf \textcolor{red}{ \( \mathbf{ \times } \) } }\\
\texttt{l}  & lake\_0.030\_May\_10\_2010 & 7x7 & Size sensitivity &{\bf \textcolor{green}{ \( \mathbf{ \surd } \) }}\\
\texttt{xxl}  & lake\_0.020\_Mar\_30\_2010 & 29x29 & Size sensitivity & {\bf \textcolor{green}{ \( \mathbf{ \surd } \) }}\\
\texttt{rough} & lake\_0.006\_Mar\_12\_2010 & 11x11 & \citet{gar92} lake-surface roughness & {\bf \textcolor{green}{ \( \mathbf{ \surd } \) }}\\
\texttt{lo\_sun}  & lake\_0.023\_Apr\_16\_2010 & 11x11 & 0.75 x present Solar luminosity  & {\bf \textcolor{green}{ \( \mathbf{ \surd } \) }}\\
\texttt{hires}  & lake\_0.004\_Feb\_10\_2010 & `11x11' & Horizontal resolution check & {\bf \textcolor{green}{ \( \mathbf{ \surd } \) } }\\
\texttt{winter}  & lake\_0.000\_May\_1\_2010 & 11x11 & Seasonal sensitivity (L$_s$=90$^\circ$) & {\bf \textcolor{green}{ \( \mathbf{ \surd } \) }} \\
\texttt{midlat}  & lake\_0.000\_Jun\_25\_2010 & 11x11 & Latitude sensitivity (latitude $\sim$ 45$^\circ$) & {\bf \textcolor{green}{ \( \mathbf{ \surd } \)} }\\
\texttt{hipress}  & lake\_0.033\_Jul\_22\_2010 & 11x11 & 60 mbar atmospheric pressure  &{\bf \textcolor{red}{ \( \mathbf{ \times } \)} }\\
\texttt{dz\_30m}  & afternoon\_dz\_test\_30m\_Jan\_11\_2010 & 11x11 & Vertical resolution check\\
\texttt{dz\_10m}  & afternoon\_dz\_test\_10m\_Jan\_11\_2010 & 11x11 &   Vertical resolution check\\
\texttt{dz\_3m}  & afternoon\_dz\_test\_3m\_Jan\_11\_2010 & 11x11 & Vertical resolution check \\
\texttt{dz\_1m}  & afternoon\_dz\_test\_1m\_Jan\_11\_2010 & 11x11 &   Vertical resolution check\\ [1ex]
%
\hline
\end{tabular}}
\label{table:nonlin}
\end{table}

 \begin{table}[ht]
\caption{Vapor fate (end of sol 6). Units are Mt (10$^6$ metric tons). Italicized \textit{\texttt{dry}} run
is subtracted from all other runs. See also Figure \ref{TERNARYFATE}.}
\centering
{\small
\begin{tabular}{l r r r r r r r}
\hline\hline
Run & Evap. rate (mm/hr) & Water in atm. & Total atm .(\%) & Snow in lake & Snow beyond lake & Total snow (\%) \\ [0.5ex]
\hline
\textit{\texttt{dry}} &  \textit{0} & \textit{420} & \textit{100\%} & \textit{0} & \textit{0} & \textit{0\%}\\
\texttt{ref (sol 6)} &  2.5 & 334 & 26\% & 234 & 741 & 74\%\\
\texttt{ref (sol 12)} & 2.5 & 316 & 11\% & 543 &  1942 & 89\% \\
\texttt{line} & 3.7 & 1420 & 12\% & 2701 & 7355 & 88\% \\
\texttt{s} &  16.2 & 26 & 38\% & 0.8 & 42.7 & 62\% \\
\texttt{m} & 3.6 & 68 & 49\% & 12 & 59.1 & 51\%\\
\texttt{l} & 2.5 & 228 & 44\%& 84 & 204 & 56\% \\
\texttt{xxl} & 2.9 & 899 & 9\% & 5030 & 4503 & 91\% \\
\texttt{rough} & 2.6 & 381 & 27\% & 250 & 793 & 73\%\\
\texttt{lo\_sun} & 2.8 & 176 & 12\% & 484 & 820 & 88\% \\
\texttt{hipress} & 3.4 & 175 & 22\% & 484 & 820 & 78\%\\ [1ex]
\hline
\end{tabular}}
\label{table:nonlin}
\end{table}

\newpage
\pagebreak

 \begin{figure}
 \noindent\includegraphics[clip=true,trim=30mm 30mm 50mm 140mm]{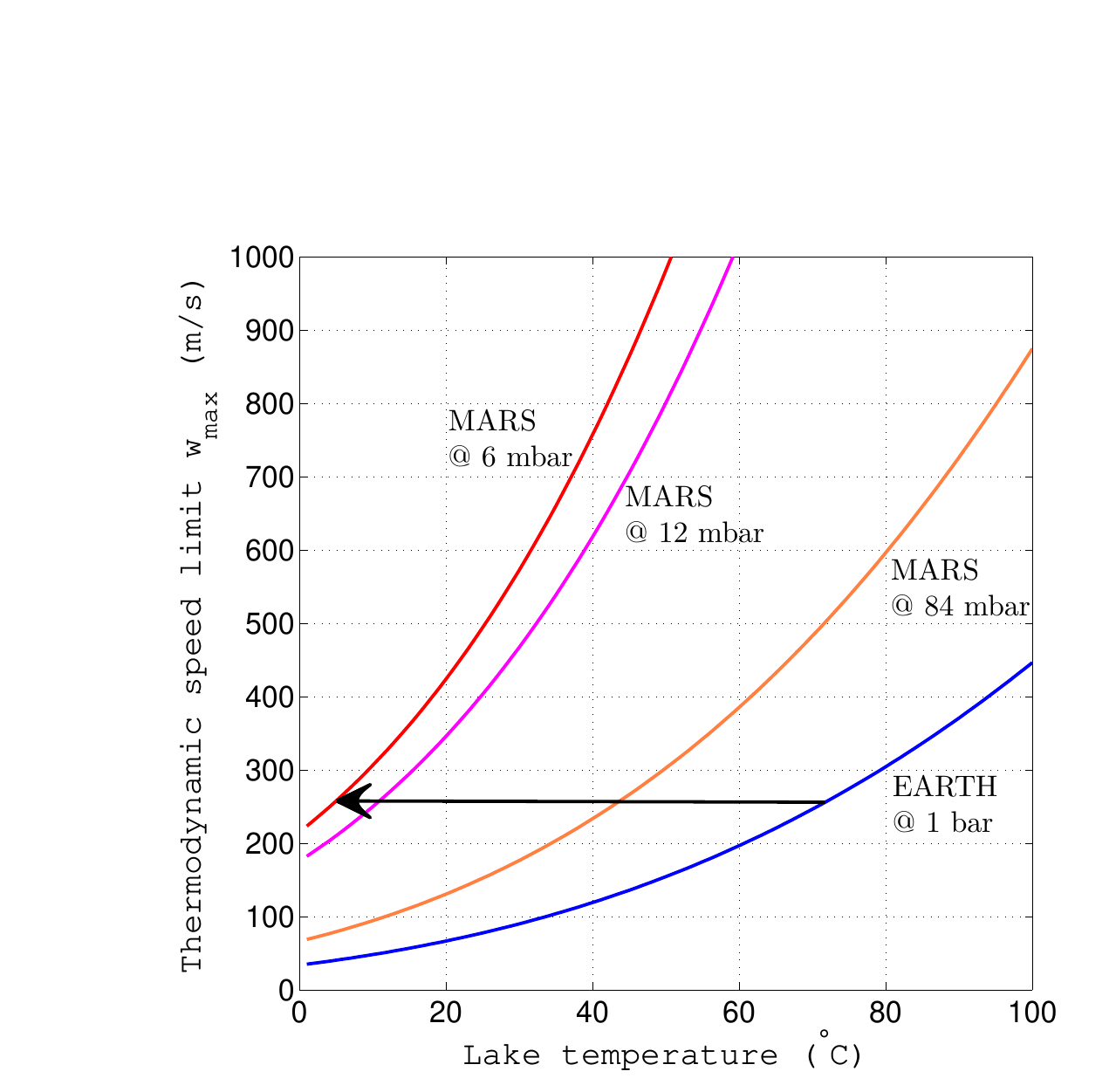}
 \end{figure}

\newpage
\clearpage
\pagebreak

 \begin{figure}
 \caption{\label{EQUIVTEMP} To show dependence of lake-driven convection on lake temperature and atmospheric pressure. Moist convection is enhanced on low-pressure Mars relative to 1-bar Earth because more buoyancy is produced for a given sea-surface temperature or lake-surface temperature. One measure of the strength of convection is the maximum vertical velocity of an updraft $w_{max}$ ( = $\sqrt{2 CAPE}$, where CAPE is Convective Available Potential Energy), the so-called thermodynamic speed limit \citep{mar10}. Suppose the Level of Free Convection (LFC) to be at the surface and the Equilibrium Level (EL) at 1.5km for both planets. Suppose complete isobaric condensation and precipitation of a parcel at 50\% humidity at the surface. In an otherwise dry atmosphere, and neglecting some second-order thermodynamic corrections, $w_{max}$ is then as shown. The red curve is for today's Mars, the blue curve is for today's Earth, and the magenta curve is is the pressure used in our Mars simulations. The black arrow shows that convection above a 5$^\circ$C lake on Mars may be as vigorous as above a 72$^\circ$ lake on Earth. The orange curve corresponds to the threshhold pressure above which we suspect localized precipitation on Mars does not occur. Values used: $c_{p,Mars}$ 770 J/kg; $c_{p,Earth}$ 1003 J/kg; $R_{Earth}$ = 287 J K$^{-1}$ mol $^{-1}$; $R_{Mars}$ = 189 J K$^{-1}$ mol $^{-1}$; $g_{Mars}$ = 3.7 m s$^{-2}$; $g_{Earth}$ = 9.8 m s$^{-2}$; $L_{sub}$ = $L_{evap}$ $\sim$ 2.5 x 10$^6$ J/kg; saturation vapor pressure curve from \citep{hardy}.}
 \end{figure}

\newpage
\pagebreak

\begin{figure}
\noindent\includegraphics[clip=true,trim=0mm 20mm 0mm 0mm,width=80mm,height=80mm]{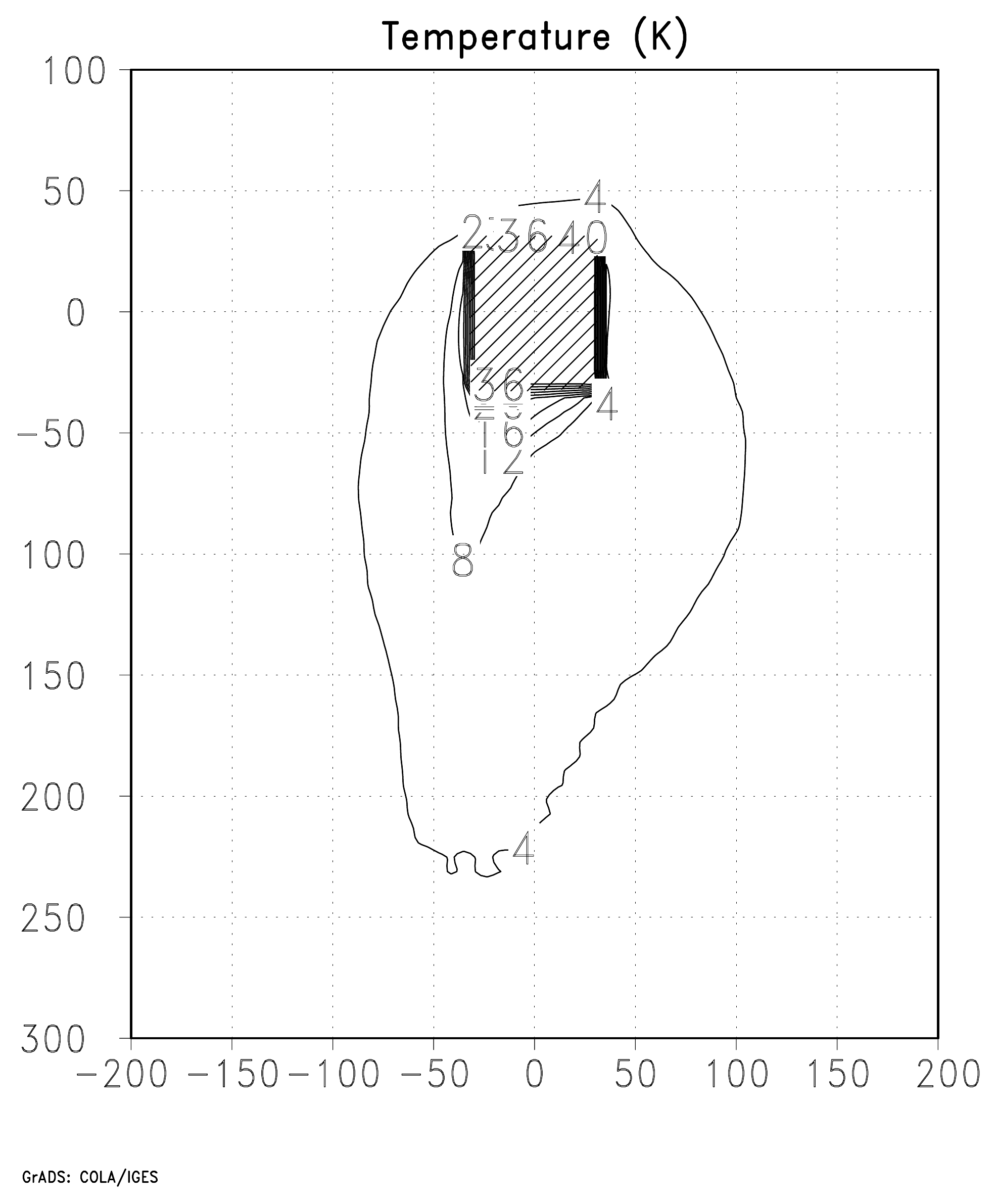}
\noindent\includegraphics[clip=true,trim=0mm 20mm 0mm 0mm,width=80mm,height=80mm]{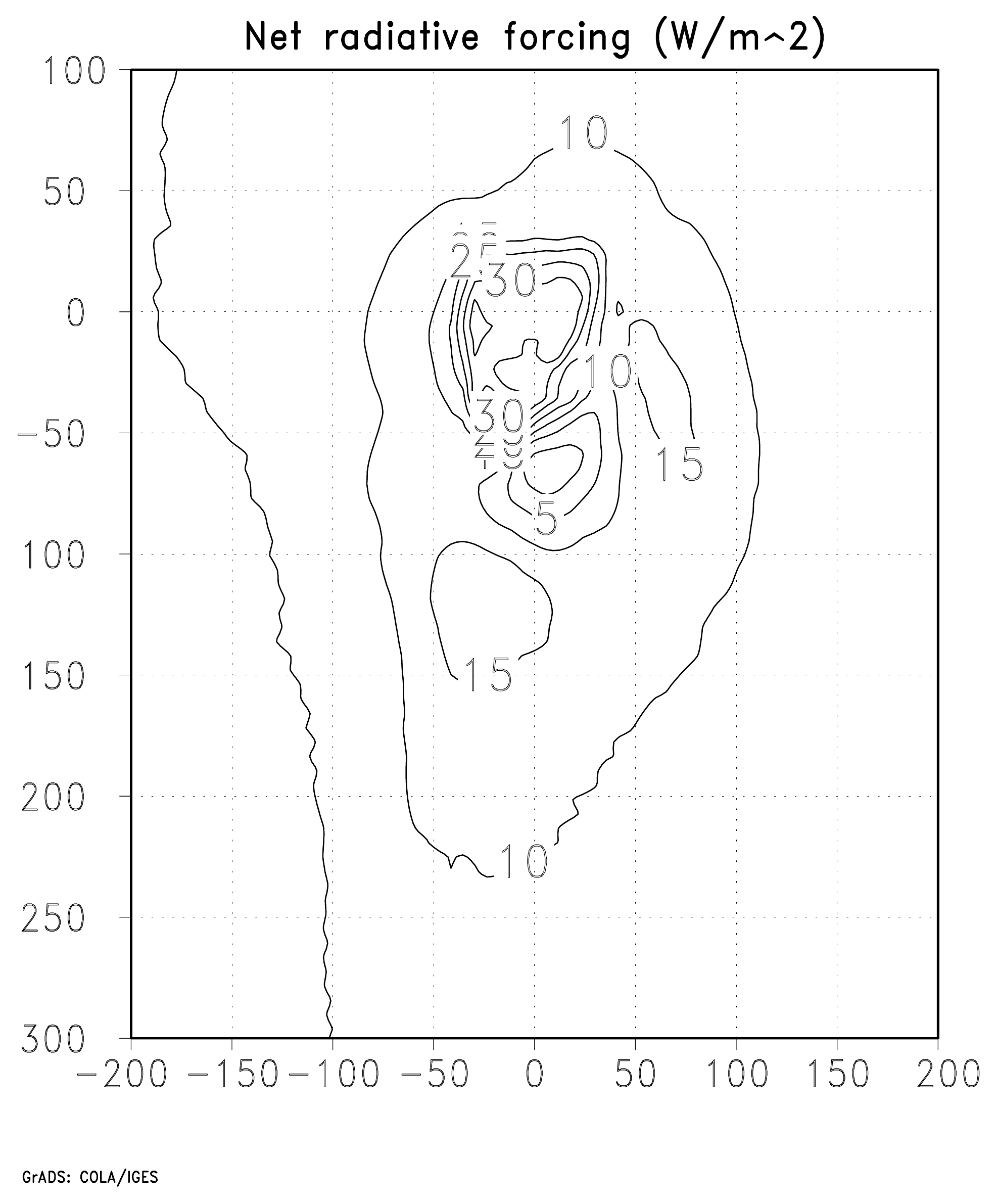}
\noindent\includegraphics[clip=true,trim=0mm 7.5mm 0mm 0mm,width=80mm,height=80mm]{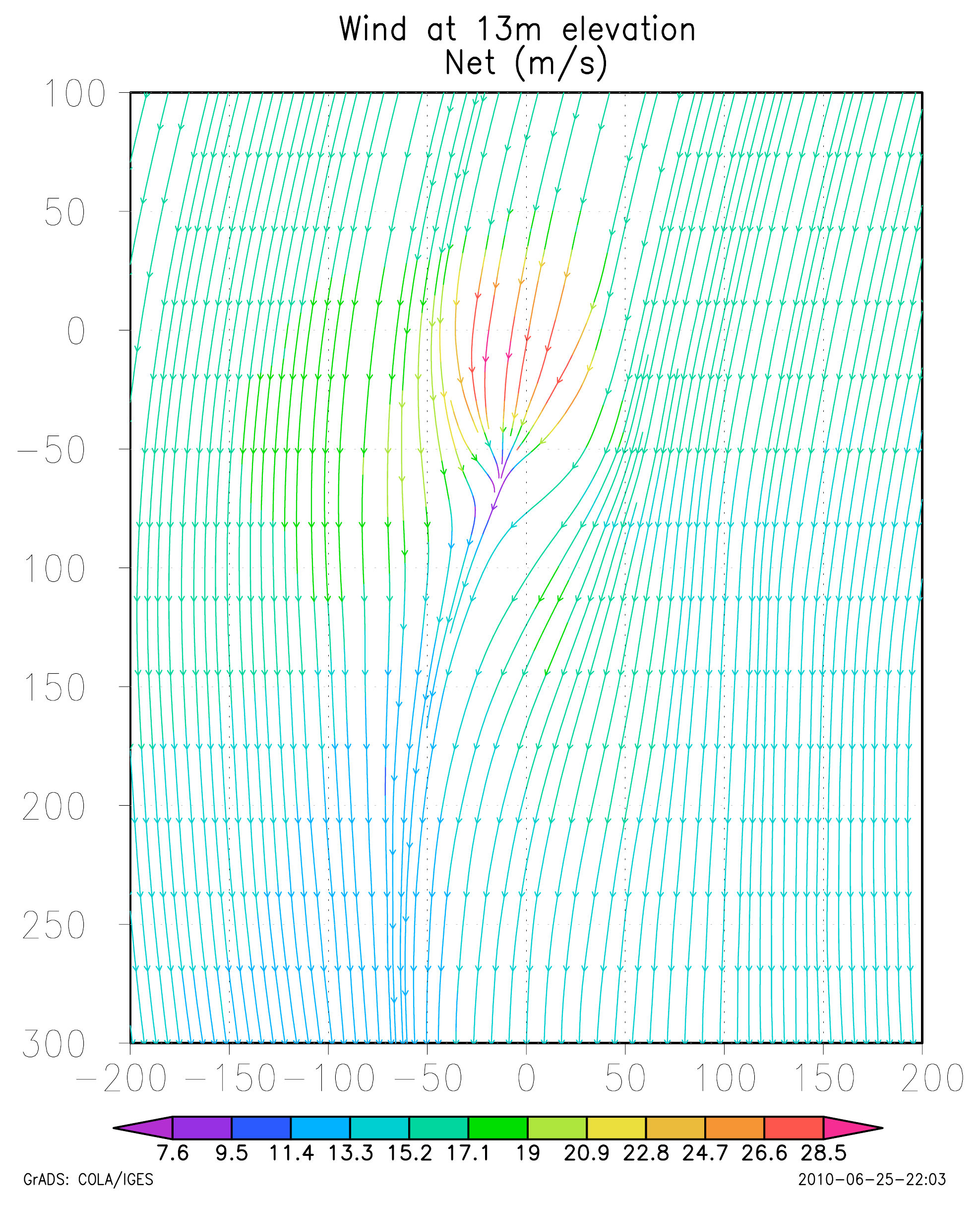}
\noindent\includegraphics[clip=true,trim=0mm 7.5mm 0mm 0mm,width=80mm,height=80mm]{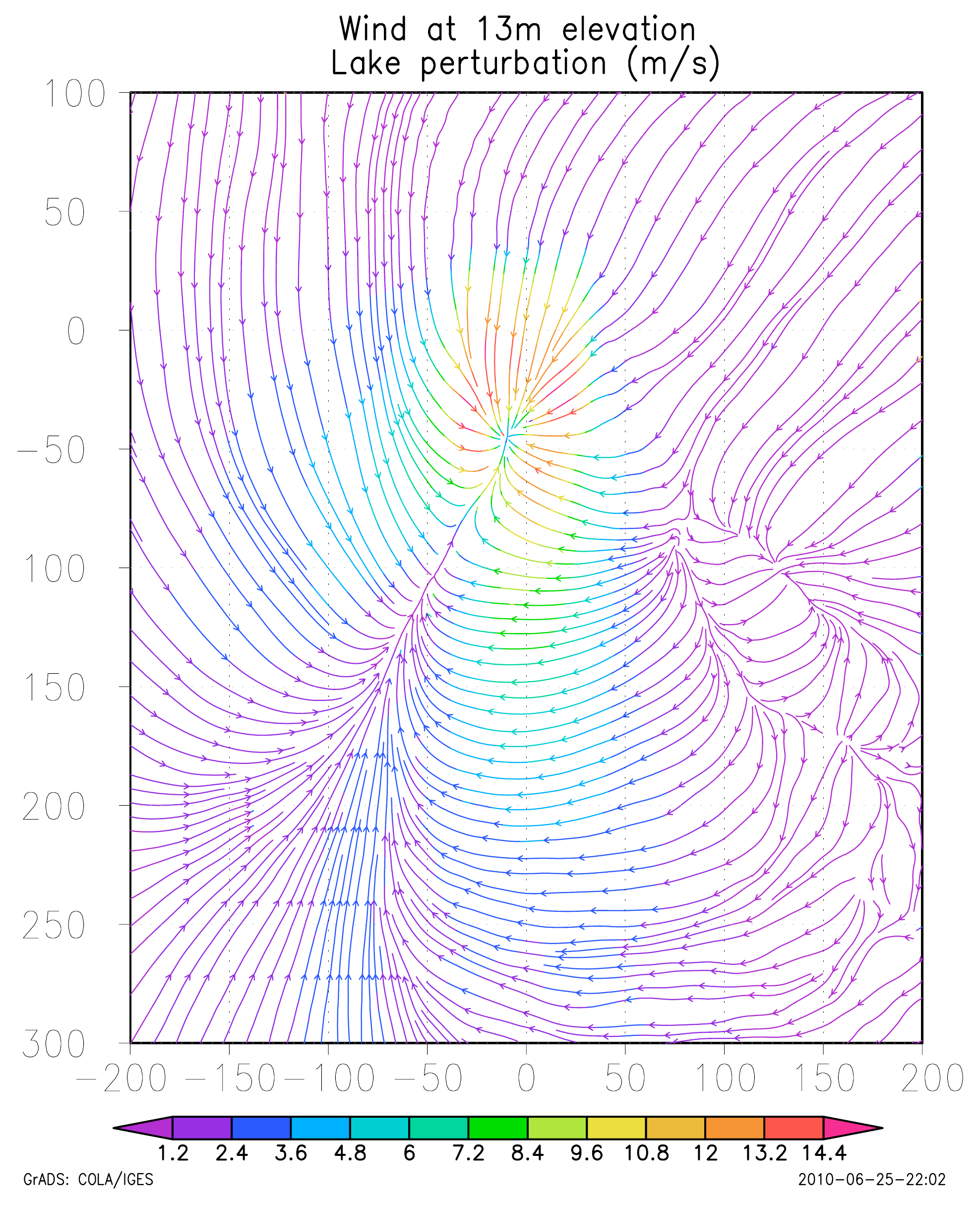}
\end{figure}


\begin{figure}
 \noindent\includegraphics[clip=true,trim=0mm 20mm 0mm 0mm,width=80mm,height=80mm]{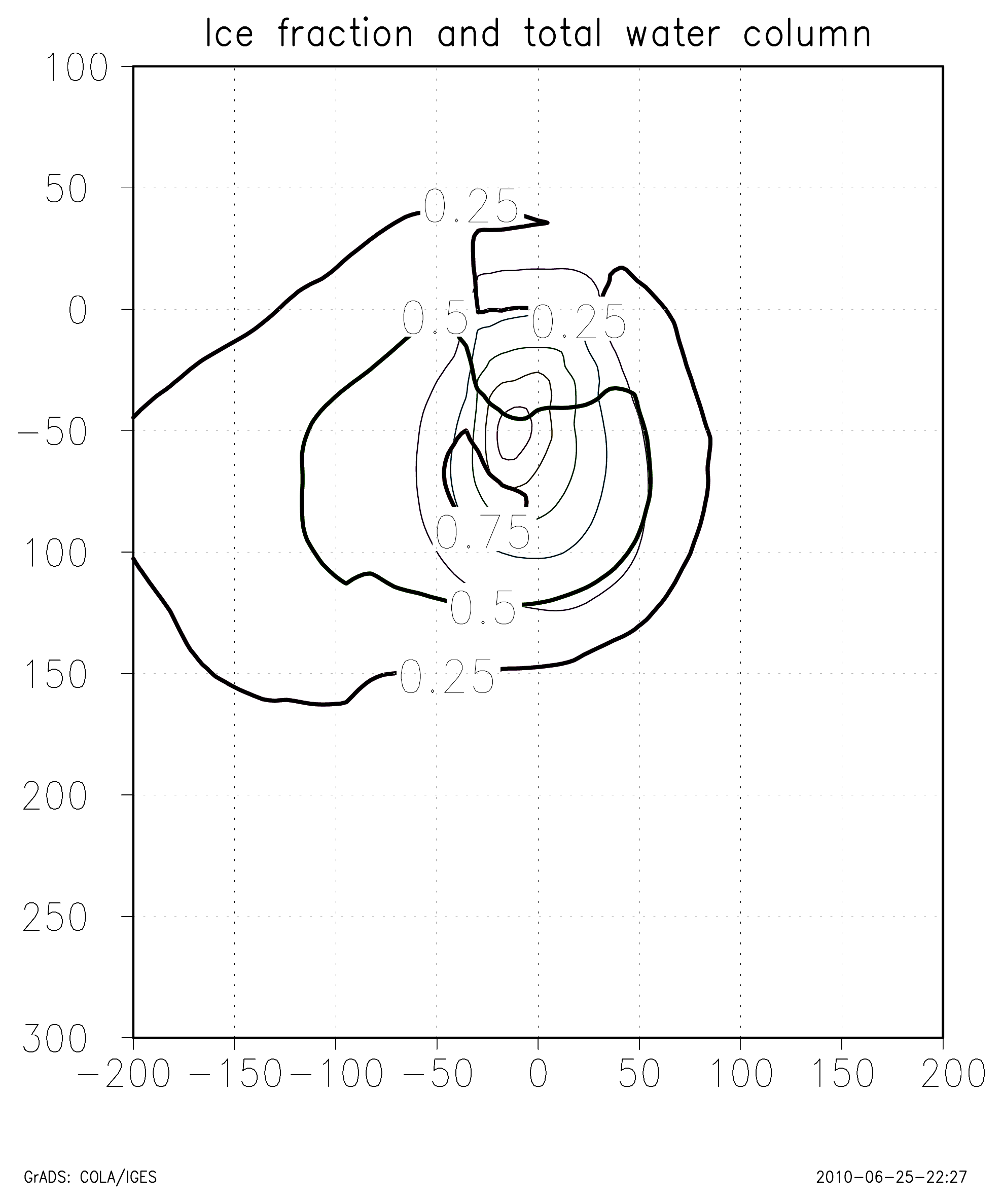}
 \caption{\label{TIMEAVERAGES} Time averaged response to the lake. The area shown is a subset of the inmost grid, and distances are in km from lake center. The shaded area in the first panel shows the extent of the lake. a) Temperature perturbation due to lake; b) radiative forcing at the surface; c) overall surface-layer wind field with lake added; b) perturbation to surface-layer wind field due to lake; e) total water column abundance (thin lines at spacing of 0.25 precipitable cm) and fractional ice abundance (thick lines).}
 \end{figure}


\begin{figure}
\noindent\includegraphics[clip=true,trim=0mm 20mm 0mm 0mm,width=80mm,height=80mm]{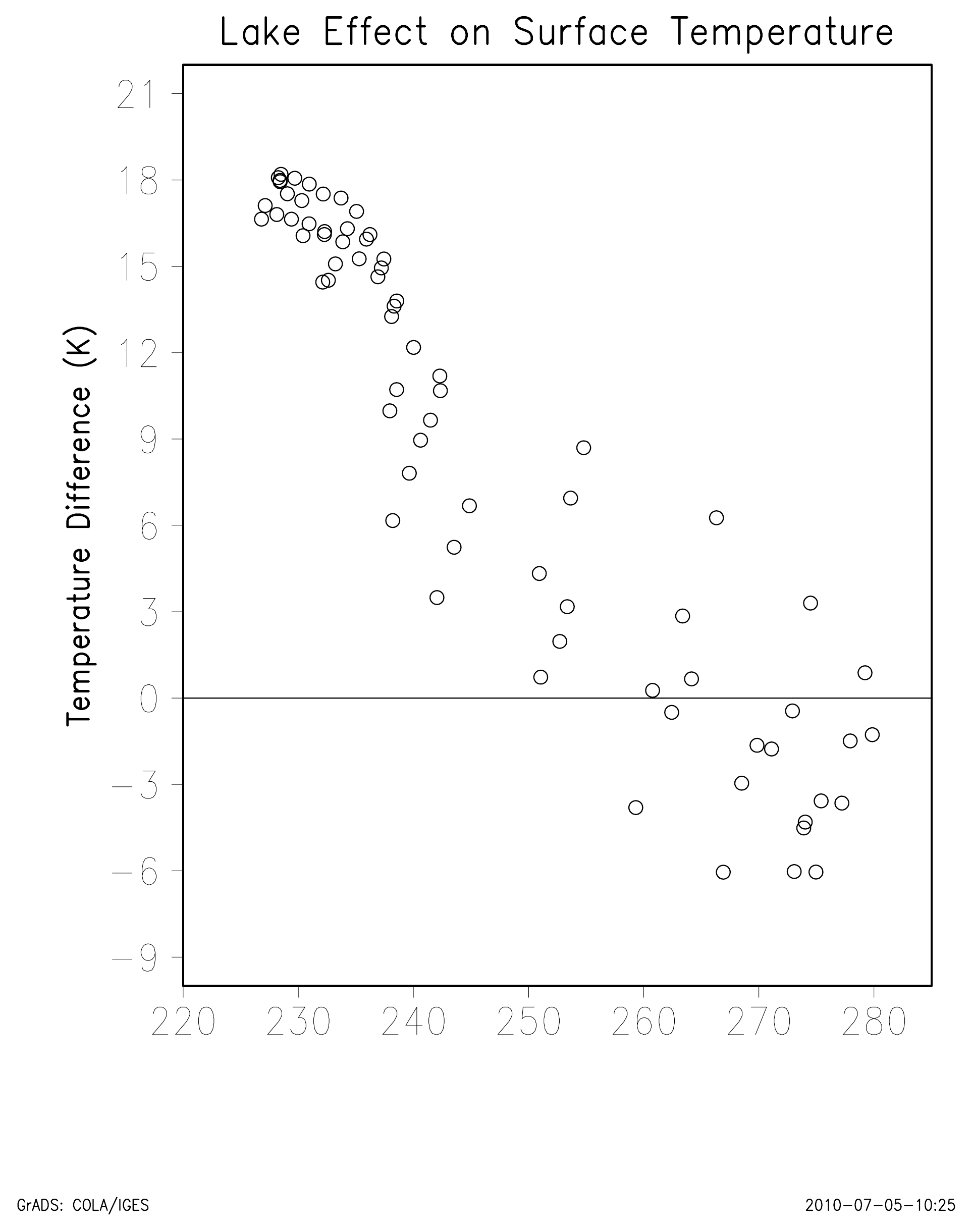}
\noindent\includegraphics[clip=true,trim=0mm 20mm 0mm 0mm,width=80mm,height=80mm]{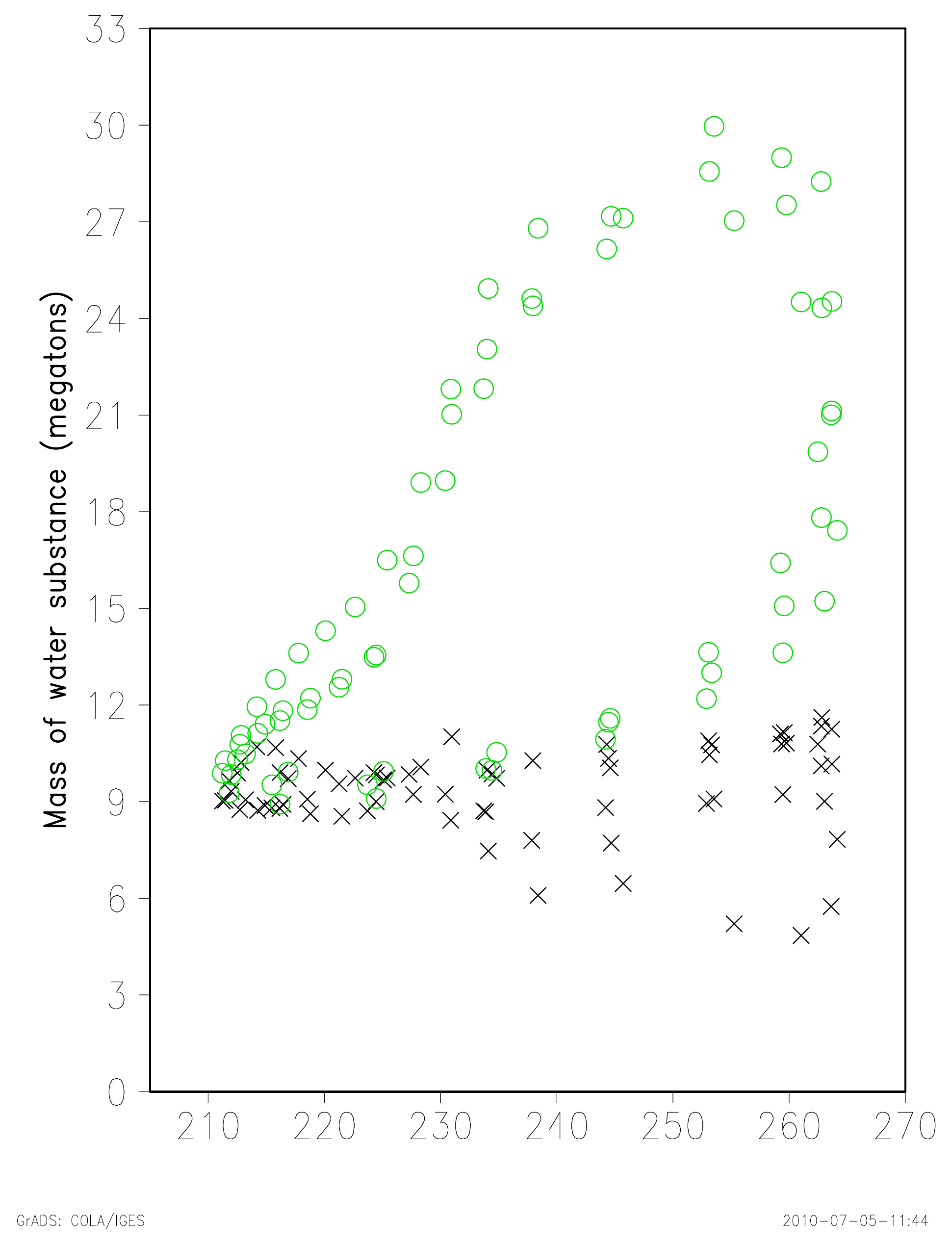}
\end{figure}


  \begin{figure}
 \caption{\label{SPATIALAVERAGES} Spatially-averaged time dependencies are highly repeatable between sols. (a) Lake perturbation to surface temperature downwind of the lake, as a function of surface temperature in K. (b) Mass of water in atmosphere in 400km side square box centered on the cloud, as a function of near-surface air temperature in K. Green circles correspond to vapor mass, and black crosses correspond to ice mass.}
 \end{figure}


  \begin{figure}
 \noindent\includegraphics[clip=true,trim=0mm 20mm 0mm 0mm,width=80mm,height=80mm]{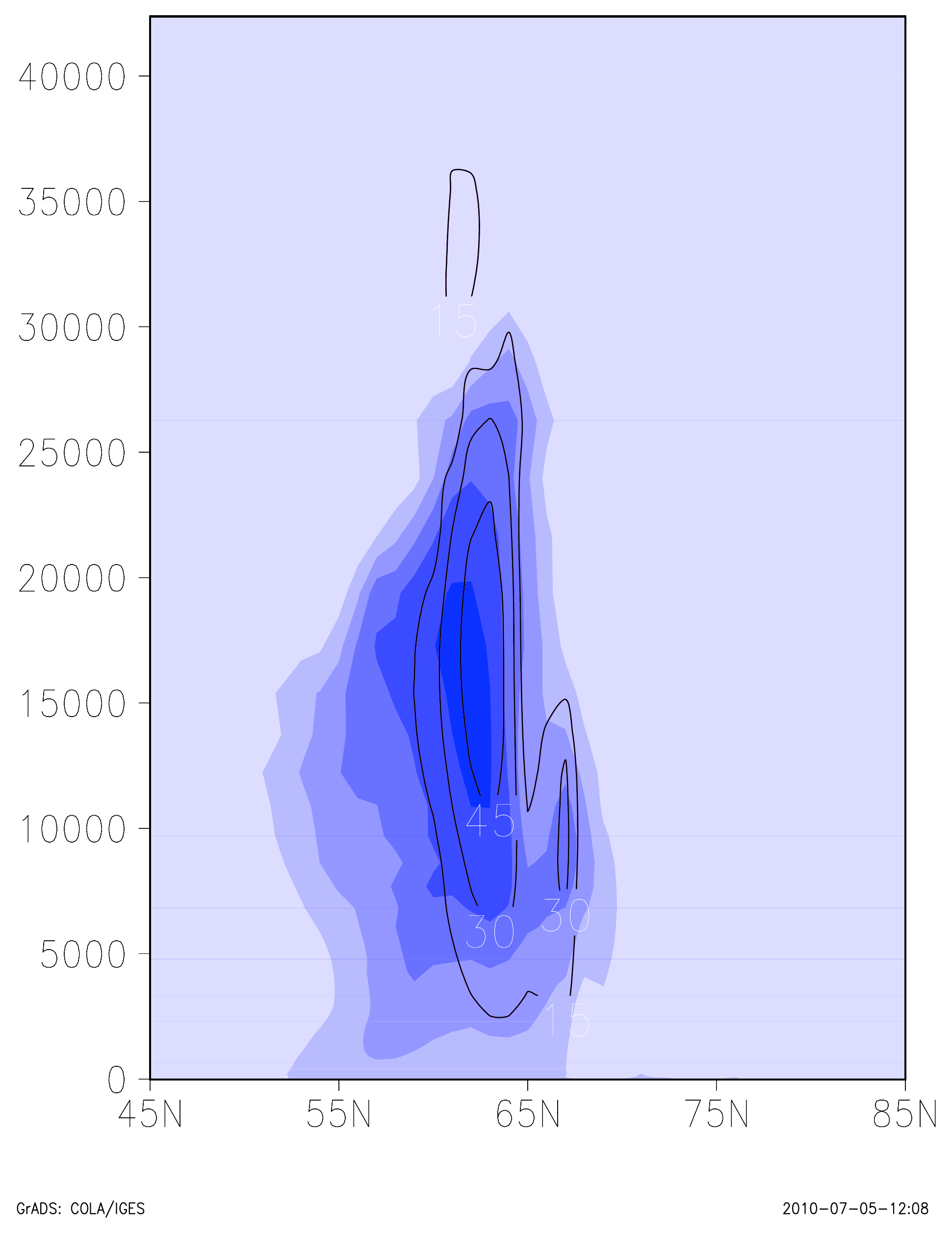}
 \caption{\label{PLUME} N-S cross section through lake storm. Blue tint corresponds to increasing water ice fraction (interval 0.002, maximum value 0.011). Labelled contours correspond to bulk vertical velocity in m s$^{-1}$. y-axis is vertical distance in m. x-axis is horizontal distance in simulation units: 10 simulation units = 59 km. Lake extends from 69N to 79N on this scale.}
 \end{figure}


\begin{figure}
\noindent\includegraphics[clip=true,trim=0mm 20mm 0mm 0mm,width=80mm,height=80mm]{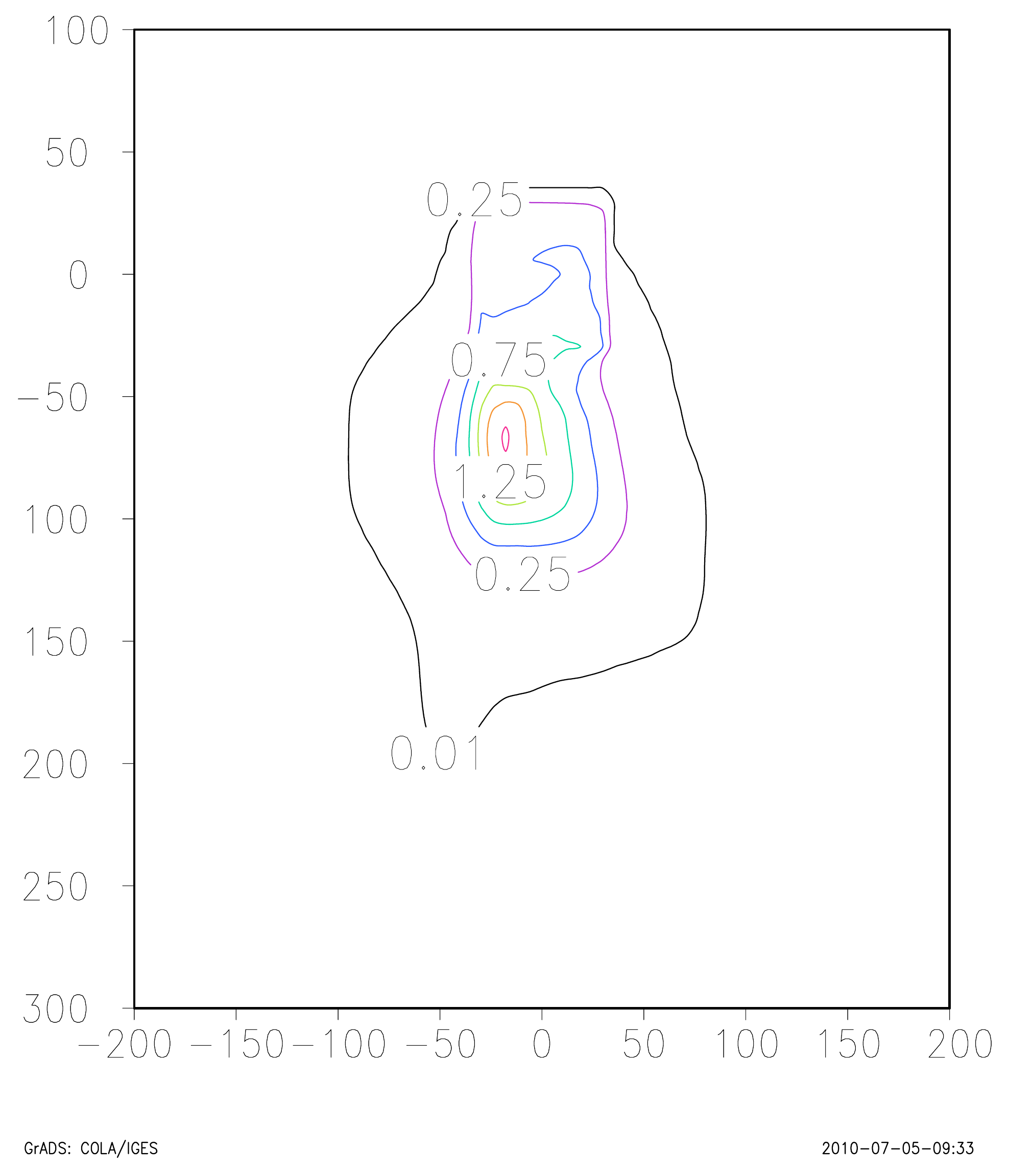}
\noindent\includegraphics[clip=true,trim=0mm 20mm 0mm 0mm,width=80mm,height=80mm]{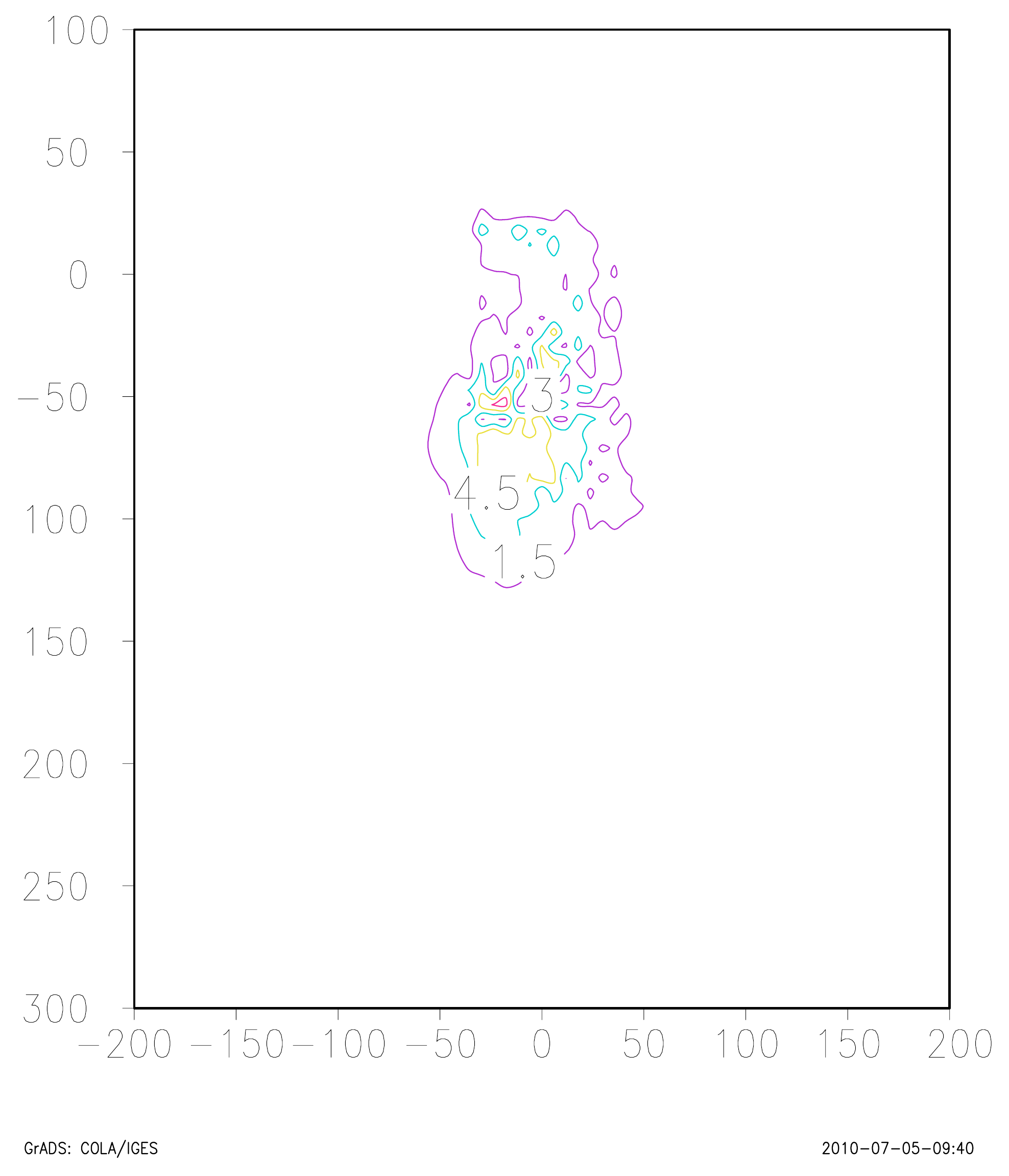}
\noindent\includegraphics[clip=true,trim=0mm 20mm 0mm 0mm,width=80mm,height=80mm]{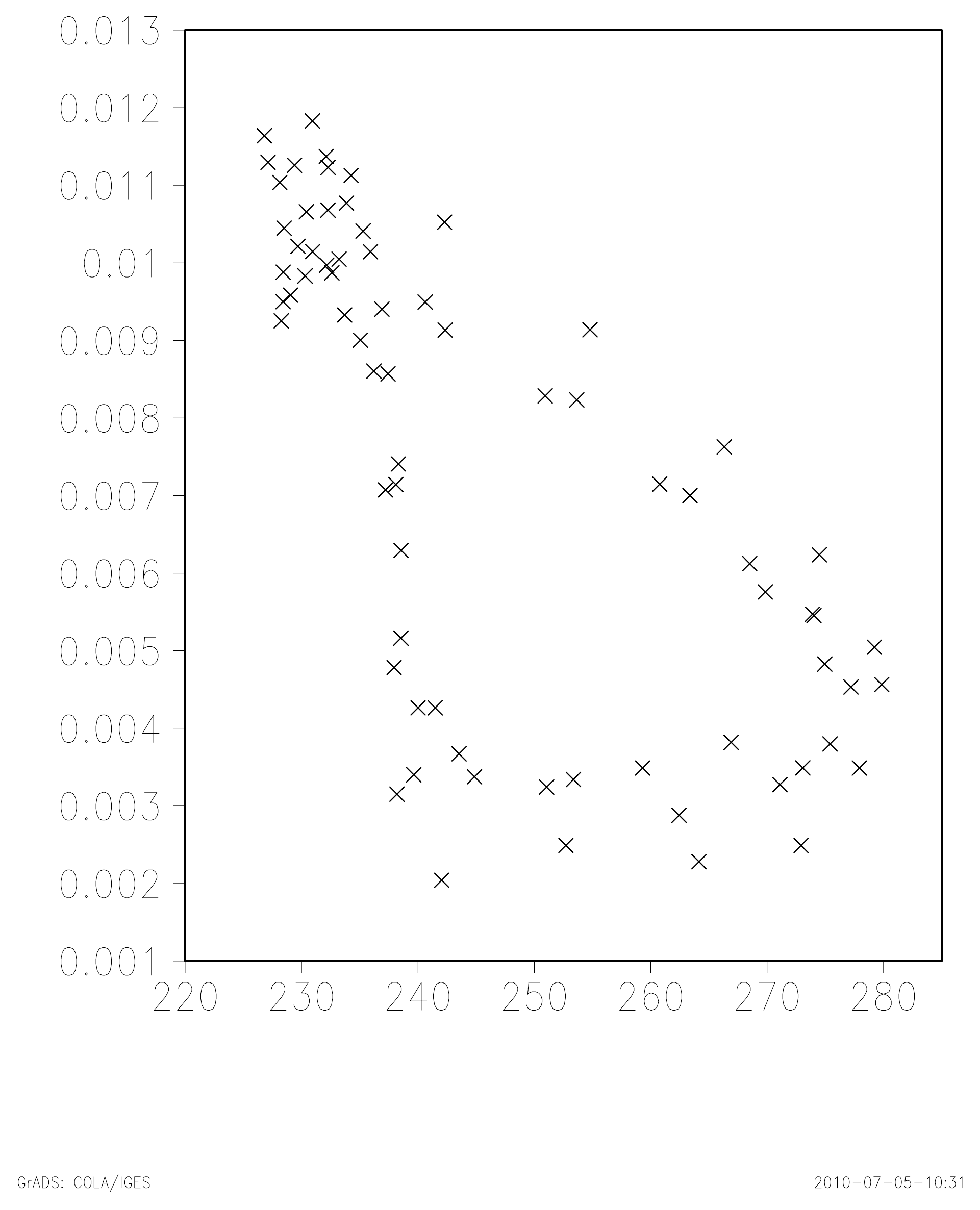}
\end{figure}

  \begin{figure}
 \caption{\label{PRECIP} Precipitation from our \texttt{ref} simulation. (a) Mean precipitation (mm/hr water equivalent) for reference simulation. Peak value is 1.5 mm/hr. (b) Maximum precipitation (mm/hr water equivalent) for reference simulation. Peak value is 6 mm/hr. (c) Diurnal cycle in spatially-averaged precipitation (mm/hr) as a function of average surface temperature (K): snow falls at a high rate during the night but is reduced following sunrise (diagonal branch), levelling out at values $\sim \frac{1}{3}$ of the nighttime peak during the late afternoon (horizontal branch). Sunset permits a rapid return to high rates of snowfall (vertical branch).}
 \end{figure}


  \begin{figure}
\noindent\includegraphics[clip=true,trim=20mm 50mm 20mm 50mm]{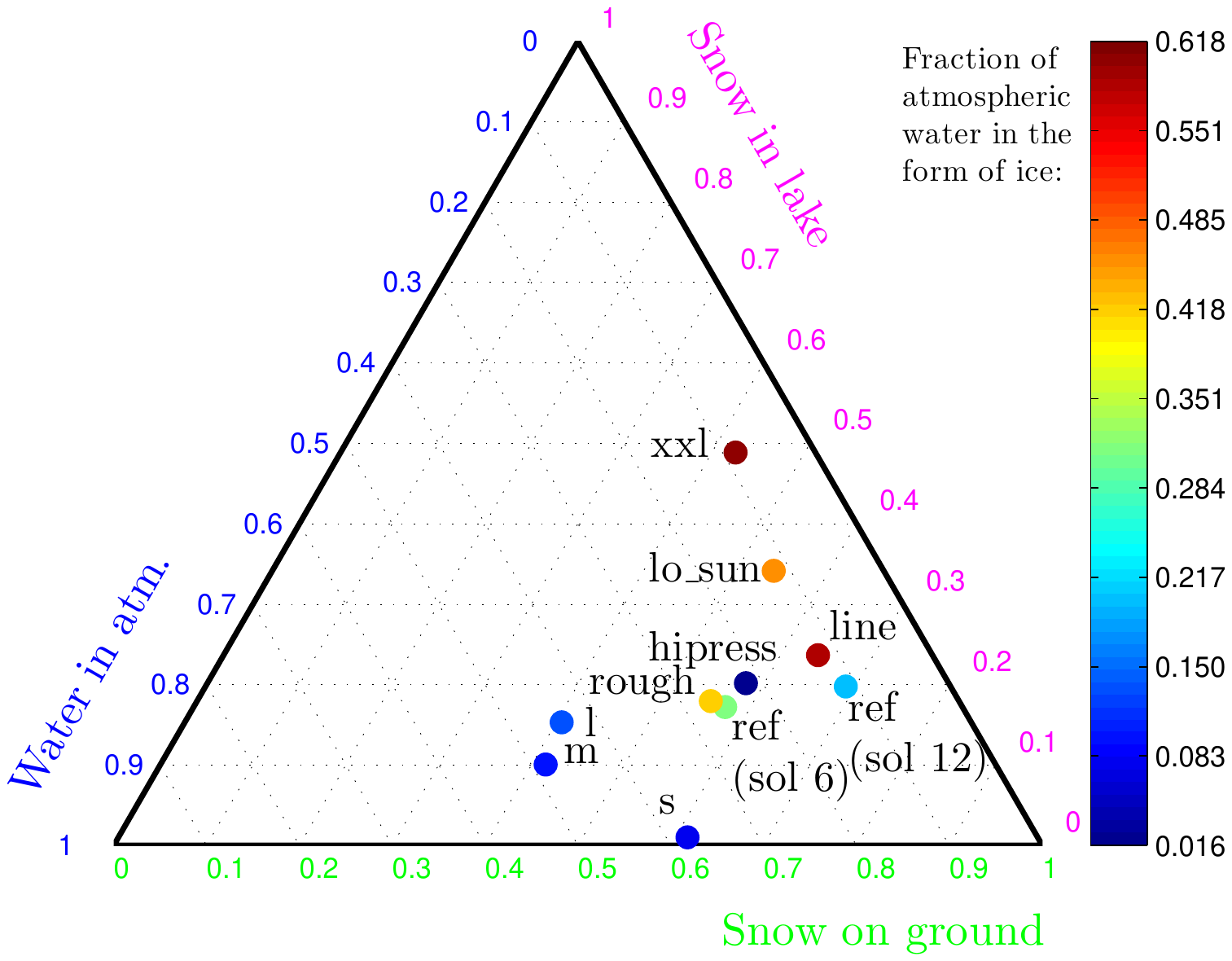}
\end{figure}


\begin{figure}
 \caption{\label{TERNARYFATE} Fate of vapor released from the lake (precipitation efficiency). Colored dots corresponds to a models' inventory of water substance after 6 sols, after substracting the inventory of a lake-free run. Color corresponds to fraction of atmospheric water in the ice phase: red is more ice-rich, blue is more vapor-rich. Only snow on ground can contribute to localized geomorphology. Water substance in atmosphere can contribute to regional and global climate change: greenhouse warming is increasingly likely as the mass of atmospheric water increases.}
 \end{figure}
\newpage
\pagebreak
\clearpage

   \begin{figure}
 \noindent\includegraphics[clip=true,trim=20mm 50mm 0mm 70mm]{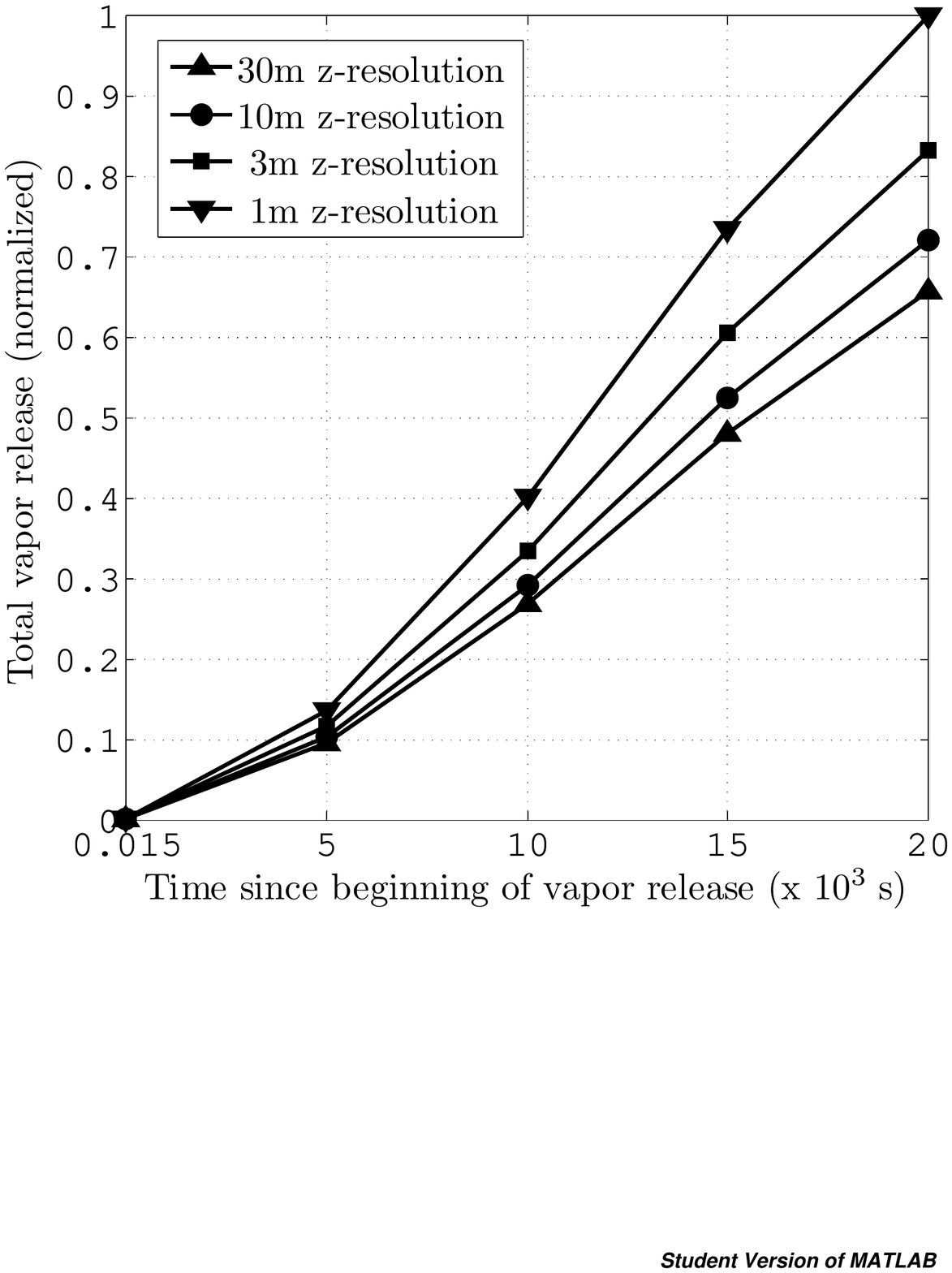}
 \caption{\label{VERTRES} Outcome of vertical resolution sensitivity test.}
 \end{figure}


  \begin{figure}
\noindent\includegraphics[clip=true,trim=20mm 0mm 0mm 110mm]{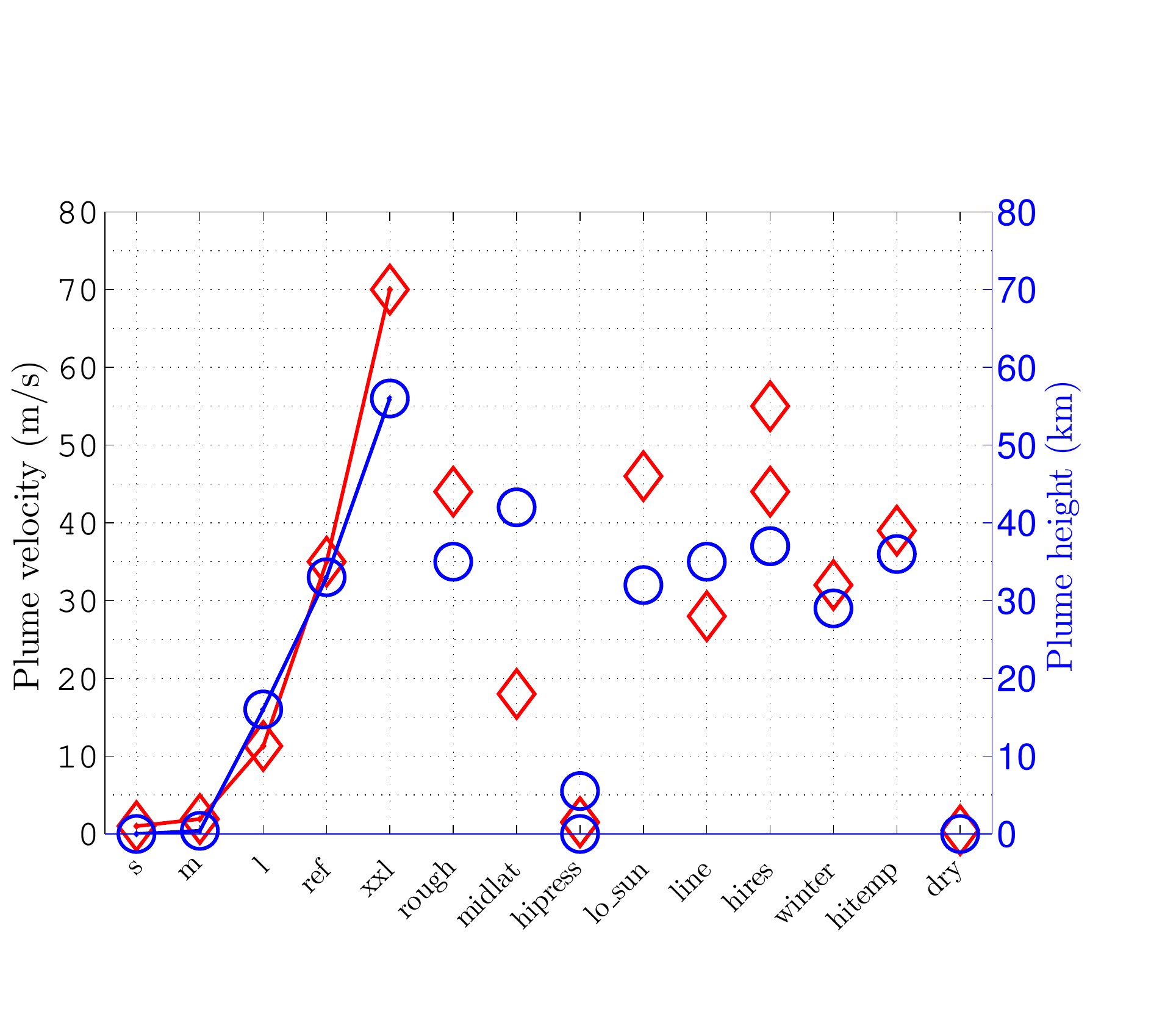}
\end{figure}


\begin{figure}
 \caption{\label{SIZETEST} Intensity of moist convection for runs listed in Table 1. Red diamonds correspond to peak time-averaged plume velocity, blue circles correspond to plume height. Lines connect the run of 5 simulations in which only size was varied. Plume height is defined as the altitude at which the plumes' spatial peak in time-averaged ice mixing ratio falls below 10$^{-3}$. The two blue circles for \texttt{hipress} correspond to ice mixing ratios of 10$^{-3}$ (lower circle) and 10$^{-4}$ (upper circle). The two red diamonds for \texttt{hires} correspond to peak velocities at 2km resolution (upper diamond) and peak velocities smoothed to 5.9 km resolution (lower diamond).}
 \end{figure}


  \begin{figure}
\noindent\includegraphics[clip=true,trim=20mm 20mm 0mm 125mm]{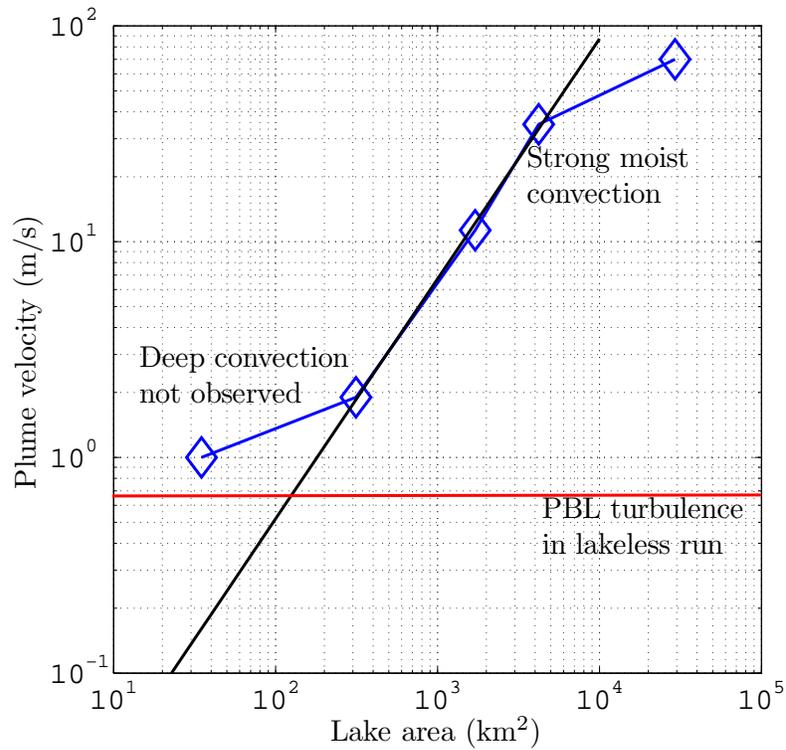}
 \caption{\label{POWERLAW} Dependence of plume convective intensity on lake size. The black line is a best-fit power law (slope = 1.11) to the three central points. It intersects the lakeless-run Planetary Boundary Layer (PBL) turbulence at a lake size of O(10$^2$)km$^2$; below this lake size, we would not expect to recognize a plume. Explanations of the similarity of the power-law exponent to 1, and the deviation of the largest lake from this slope, are provided in the text.}
 \end{figure}


  \begin{figure}
\noindent\includegraphics[width=140mm,clip=true,trim=40mm 0mm 40mm 135mm]{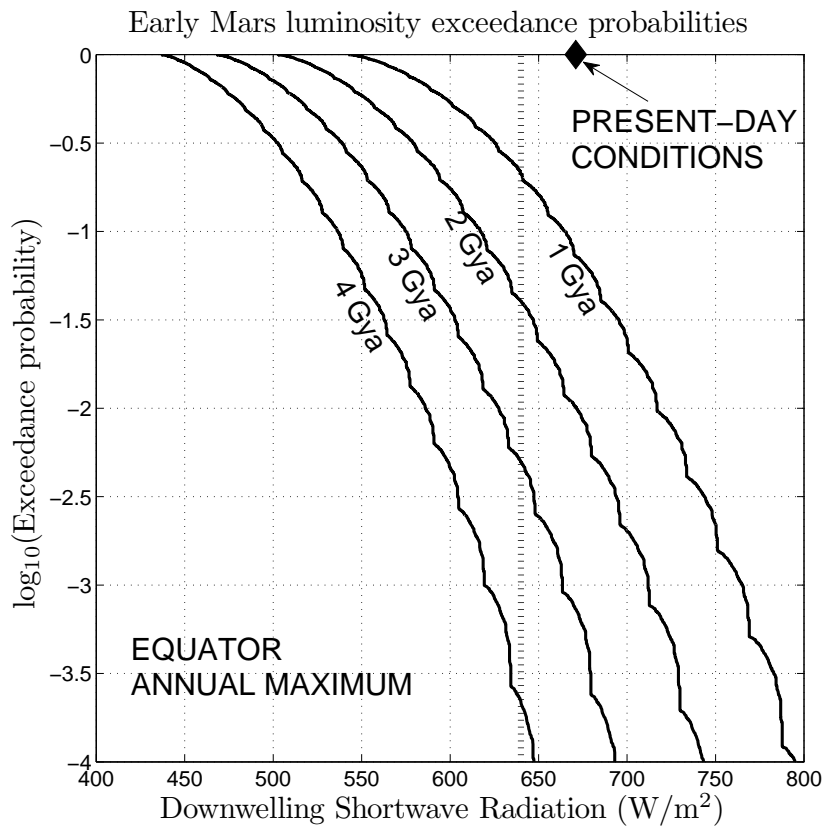}
 \caption{\label{EXCEEDANCE} Exceedance probabilities for annual maximum sunlight at Mars' equator. Vertical dashed line is 640W/m$^2$, the minimum for melting (see text). The small wiggles in the solid curves are interpolation artifacts. Snow instantaneously emplaced at Mars' equator with the current orbital conditions and solar luminosity would melt. For progressively earlier times (reduced solar luminosity), the probability of melting decreases.}
 \end{figure}


  \begin{figure}
 \noindent\includegraphics[width=140mm,clip=true,trim=40mm 0mm 40mm 135mm]{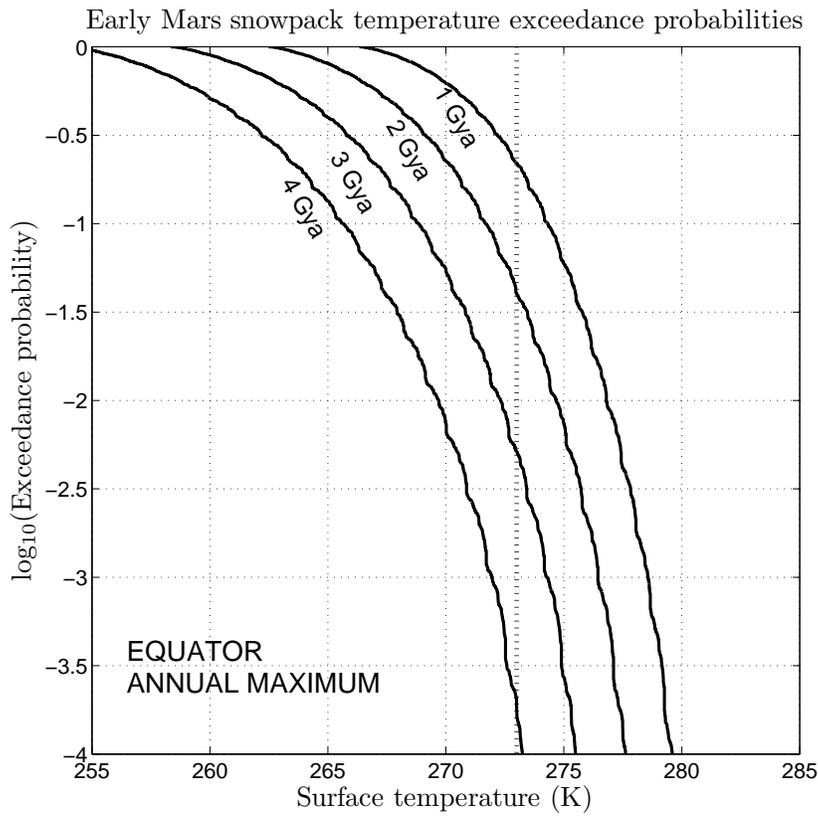}
 \caption{\label{CDFS} Exceedance probabilities for annual peak temperature of snowpack at Mars' equator, from our 1D column model. A dust-like albedo (0.28) is assumed. Vertical dashed line corresponds to the melting point. The small wiggles in the solid curves are interpolation artifacts. For progressively earlier times (reduced solar luminosity), the probability of melting decreases. }
 \end{figure}


%
%

 \begin{figure}
 \noindent\includegraphics[width=85mm,clip=true,trim=55mm 70mm 35mm 70mm]{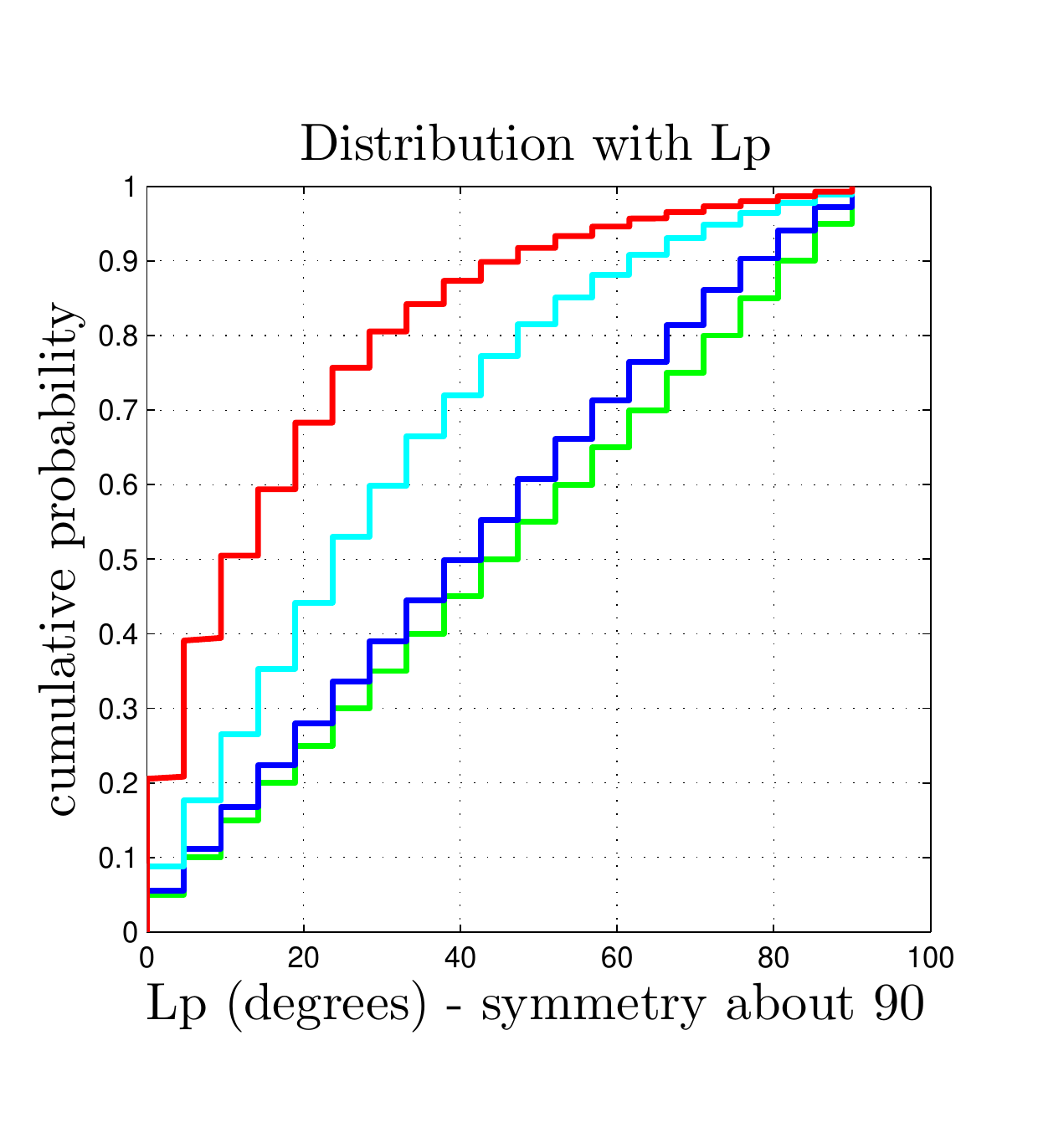}
 \noindent\includegraphics[width=85mm,clip=true,trim=55mm 70mm 35mm 70mm]{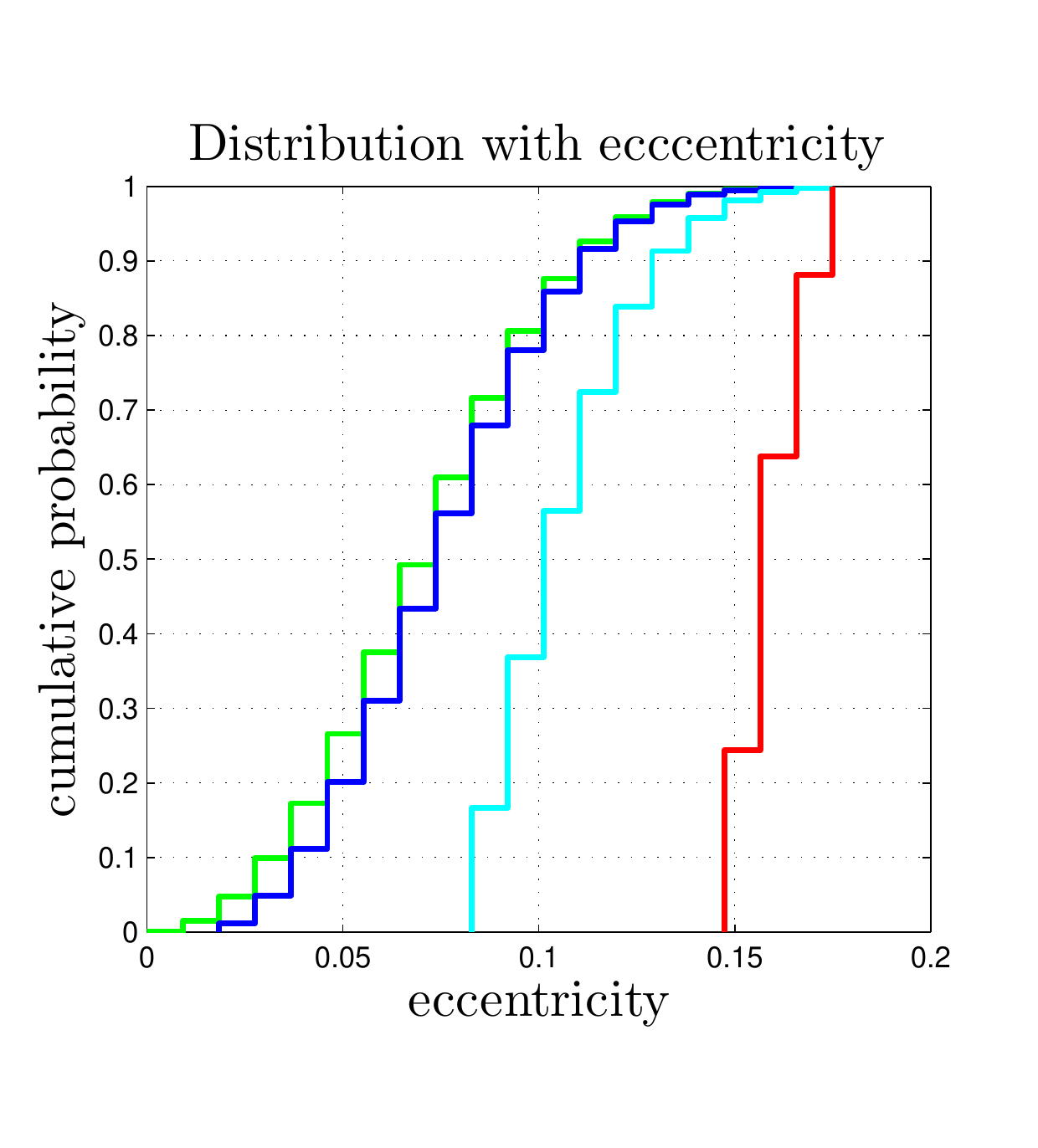}
 \noindent\includegraphics[width=85mm,clip=true,trim=55mm 70mm 35mm 70mm]{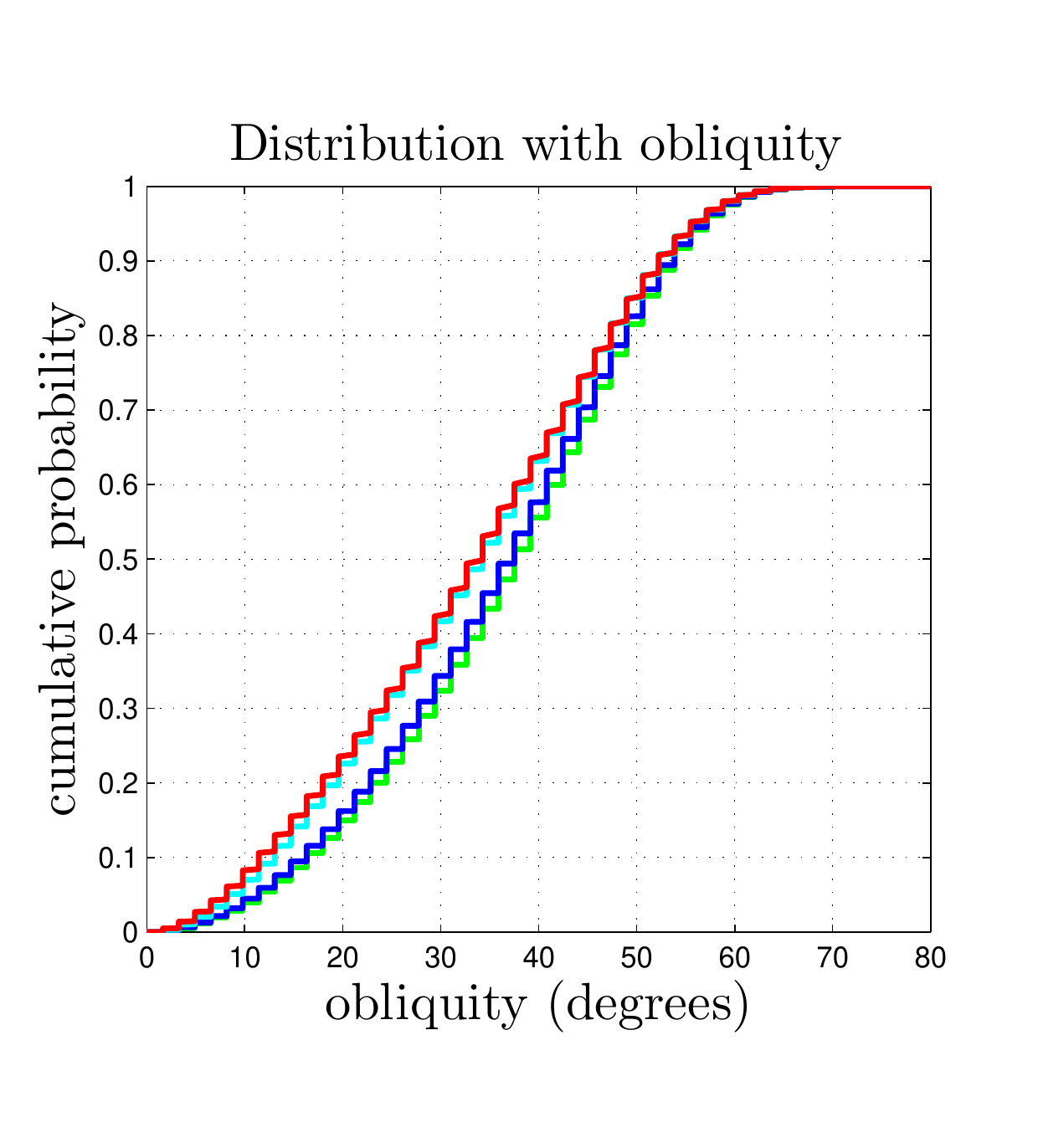}
 \end{figure}

    \begin{figure}
 \caption{\label{MONTECARLO} Self-normalized probability density functions for the subset of sampled orbital conditions that produce melt. Albedo = 0.28, age = 1.0 Gya (other albedos and ages have very similar pdf shapes, although maximum temperatures differ). The cyan lines corresponds to 273K (melting), and the other colors are green (all temperatures), blue (268K), and red (278K).}
 \end{figure}



 \begin{figure}
 \noindent\includegraphics[width=85mm,clip=true,trim=60mm 0mm 65mm 160mm]{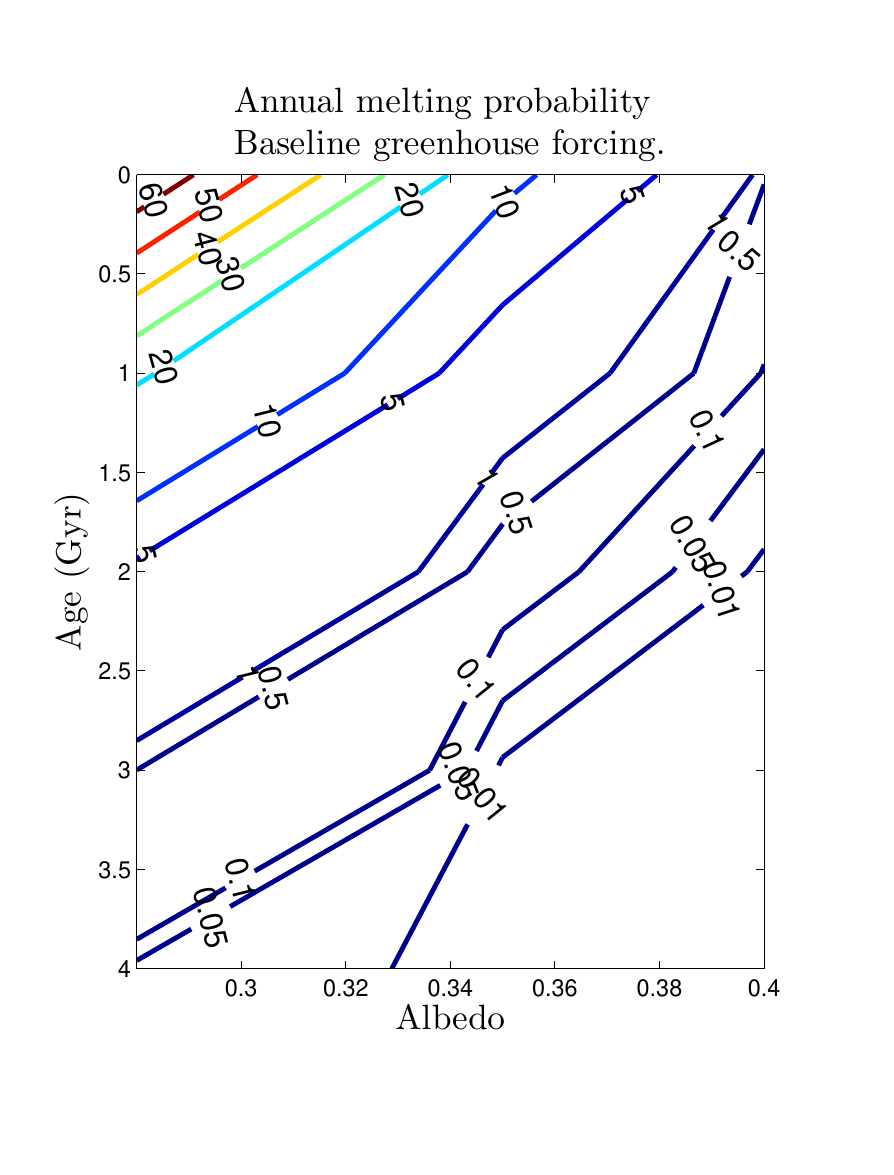}
 \noindent\includegraphics[width=85mm,clip=true,trim=60mm 0mm 65mm 160mm]{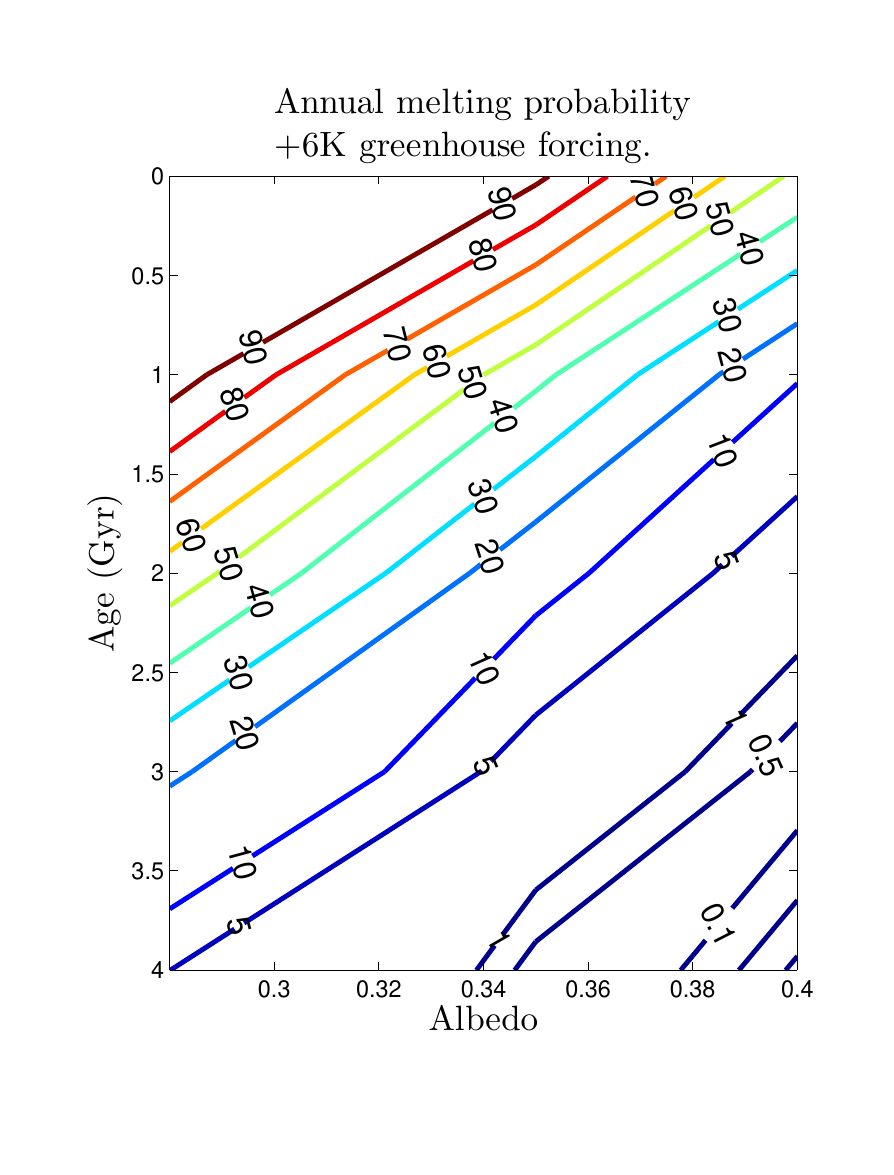}
 \end{figure}

  \begin{figure}
 \caption{\label{ALBEDOLUMINOSITY} Contoured probability (percent) of melting of rapidly-emplaced equatorial precipitation as a function of albedo and solar luminosity. In the case of impact-induced precipitation, snow falling on hot ejecta will melt regardless of orbital conditions.
}
 \end{figure}

%
%

  \begin{figure}
 \includegraphics[width=140mm,angle=90]{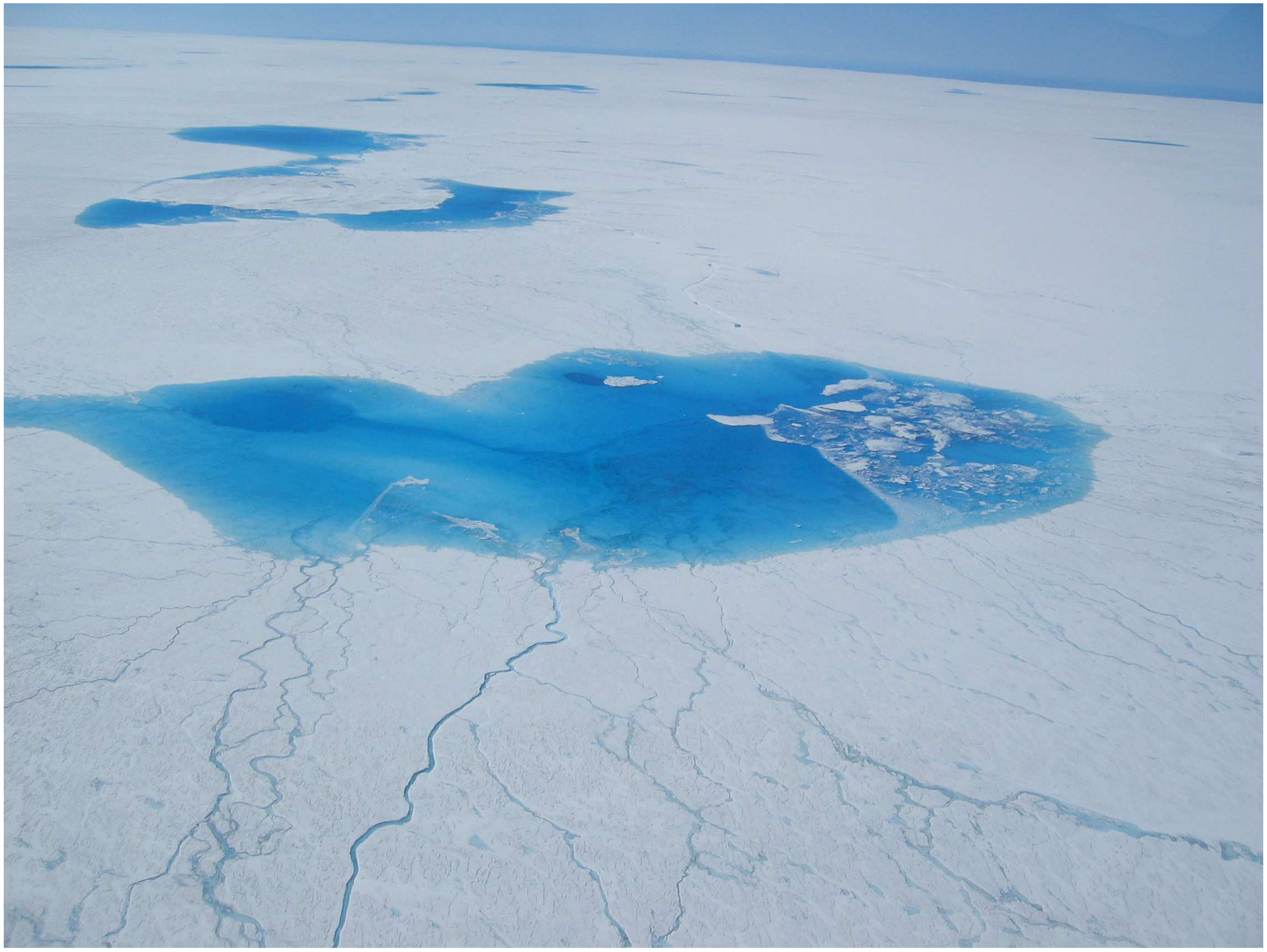}
 \caption{\label{GREENLANDPOOL} Seasonal supraglacial lakes on Greenland ice sheet.  Note absence of clouds and precipitation at 1 bar, and high drainage density of meandering, snowmelt-fed channels feeding lake. Largest lake is 4.5 km across. (Image used with kind permission of I. Joughin, U. Washington Polar Science Center).}
 \end{figure}

\begin{figure}
\noindent\includegraphics[width=170mm,clip=true,trim=210mm 450mm 160mm 620mm]{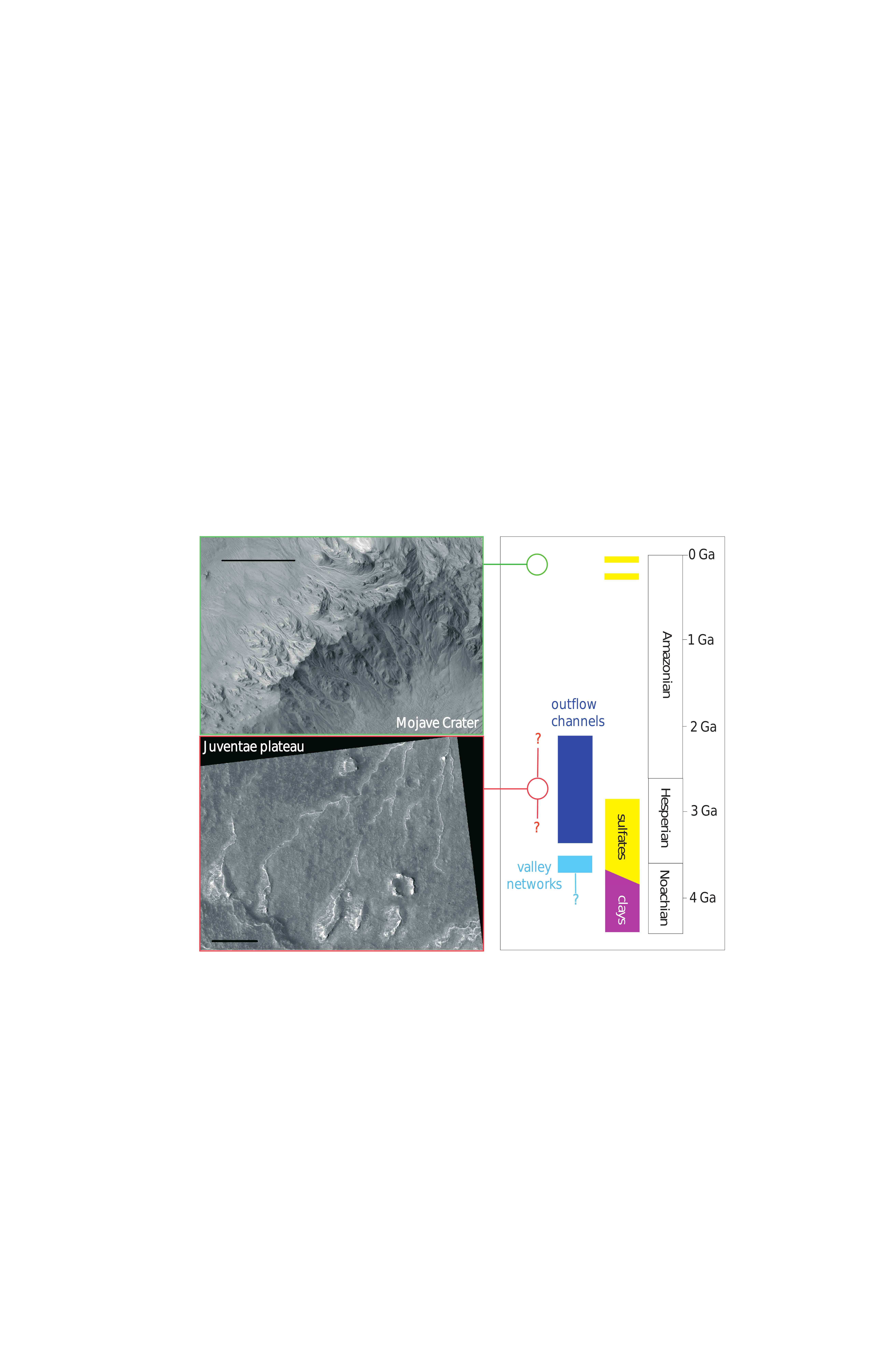}
\end{figure}
%

  \begin{figure}
 \caption{\label{MOTIVATION} Primary motivation for this study: Two geological settings where fluvial activity postdates the Late Noachian - Early Hesperian acme of channel formation, and there is a candidate vapor source nearby. Scale bars are 500 m.{\it Green box:} Sub-kilometer fans resembling alluvial fans at Mojave Crater, which may be a recent impact into icy ground (7.9N 326.6E, orthorectified HiRISE image PSP\_001481\_1875). Channels dissect both sides of 100-200m tall ridge and channel heads are found $<$ 20m from ridgeline.; {\it Red box:} Inverted streams on plateau near Juventae Chasma (4.3S 296.7E, orthorectified HiRISE image PSP\_003223\_1755). Negligible post-Noachian erosion over most of the rest of the planet indicates that such events were localized. {\it Sources for stratigraphic context:} \citet{harrecent} (absolute ages); \citet{mur09} (clay and sulfate stratigraphy) \citet{car10} (age of outflow channels); \citet{fas08date} (age of valley networks); \citet{mas10} \& \citet{man10} (existence of young sulfates); \citet{led10} (age of Juventae plateau channels).}
 \end{figure}

 \end{document}